\documentclass[aps,pra,showpacs,superscriptaddress,groupedaddress]{revtex4}  

\usepackage{graphicx}   
\usepackage{bm}             
\usepackage{amssymb}   
\usepackage{amsmath}
\usepackage{epsfig}
\usepackage{color}
\usepackage{array}
\usepackage{multirow}

\begin{document}


\title{Vibrational excitation and dissociation of deuterium molecule by electron impact}

\author{V.~Laporta}
\email{vincenzo.laporta@istp.cnr.it}
\affiliation{Istituto per la Scienza e Tecnologia dei Plasmi, CNR, 70126 Bari, Italy}

\author{R.~Agnello}
\affiliation{Ecole Polytechnique F\'{e}d\'{e}rale de Lausanne, Swiss Plasma Center, CH-1015 Lausanne,
Switzerland}

\author{G.~Fubiani}
\affiliation{LAPLACE, Universit\'{e} de Toulouse, CNRS, 31062 Toulouse, France}

\author{I.~Furno}
\affiliation{Ecole Polytechnique F\'{e}d\'{e}rale de Lausanne, Swiss Plasma Center, CH-1015 Lausanne,
Switzerland}

\author{C.~Hill}
\affiliation{Nuclear Data Section, International Atomic Energy Agency, A-1400 Vienna, Austria}

\author{D.~Reiter}
\affiliation{Institute for Laser and Plasma Physics, Heinrich-Heine-University, D-40225 D\"{u}sseldorf, Germany}

\author{F.~Taccogna}
\affiliation{Istituto per la Scienza e Tecnologia dei Plasmi, CNR, 70126 Bari, Italy}

\begin{abstract}
A theoretical investigation on electron-D$_2$ resonant collisions -- \textit{via} the low-lying and the Rydberg states of D$_2^-$ -- is presented for vibrational excitation, dissociative electron attachment and dissociative excitation processes by using the Local-Complex-Potential approach. Full sets of vibrationally resolved cross sections, involving the ground electronic state -- $\textrm{X}\,^1\Sigma^+_g$ -- and the first two electronic excited states -- $\textrm{b}\,^3\Sigma^+_u$ and $\textrm{B}\,^1\Sigma^+_u$ -- of the D$_2$ molecule, are given for fusion plasma applications in their  technologically relevant partially dissociated, detached divertor regimes. In particular, transitions between electronic excited states are also considered. Comparisons are made with cross sections present in the literature, where available.
\end{abstract}

\pacs{xxxxx}

\maketitle

\section{Introduction \label{sec:intro}}

Electron-impact vibrational excitation and dissociation of deuterium molecules are of primary importance in many fields, ranging from astrophysics and plasma discharges to nuclear fusion, including fundamental physics~\cite{Krishnakumar:2018aa}.

Molecular hydrogen H$_2$, deuterated hydrogen HD and deuterium D$_2$, the simplest and at the same time the most abundant species formed in the pregalactic gas prior to structure formation, played an important role in the cooling of the gas clouds which gave birth to the first stellar generation~\cite{1742-6596-4-1-002, doi:10.1146/annurev-astro-082812-141029}. In this regard, it has been shown that the D/H ratio in giant planets, Jupiter and Saturn, is a fundamental parameter to understand the formation of the solar system from the primitive nebula~\cite{LECLUSE19961579}. Moreover, electron-impact cross sections for HD, D$_2$ and the corresponding ions are needed to explain certain phenomena occurring in the different planetary atmospheres and their ionospheres~\cite{https://doi.org/10.1002/2013JA019022}.

Among technological applications, electron-D$_2$ collision cross sections are strictly related to the important problem of plasma interaction with the neutral gas component originating from surface recombination, from volumetric recombination channels, and/or from external gas puffing (plasma fuelling). This is of particular relevance for the so-called ``detached divertor regime" in magnetic fusion devices~\cite{Sang_2020, Verhaegh_2021}, which is  currently intensively studied in existing tokamaks and stellarators, and is also foreseen as the standard operational mode for the ITER fusion reactor under construction in Cadarache (South France). In this detached plasma mode of operation the hydrogenic plasma chemistry plays a key role for critical issues such as plasma energy and momentum dissipation, high heat flux component protection and particle (ash) removal. There molecular reaction kinetics comprises an important sub-model of the integrated multi-physics boundary plasma code systems, such as \textit{e.g.} SOLPS-ITER, which was and still is  heavily used for guiding the ITER divertor design \cite{PITTS2019100696}, and future experimental campaigns.

Powerful volumetric plasma-gas interaction mechanisms are known to be operative in fusion machine reactors, involving both resonant electron collisions -- \textit{via} resonant D$_2^-$ anion states -- and near resonant positive atomic ion (ion conversion, such as charge transfer channels) reactions. Such resonant processes are believed to be a key ingredient towards quantitative understanding of the near target divertor plasma dynamics, with their prominent effects also on dissociation degree, perhaps molecular-assisted additional volumetric recombination channels, but certainly for establishing the vibrational distribution of electronic ground molecules. Eliminating this latter from the unknown parts of fusion divertor models by a reliable cross section database can be expected to greatly improve the predictive and interpretative quality in current fusion reactor divertor models and spectroscopy interpretation tools. A recent account of the status of  including such resonant channels in fusion relevant collision radiative models is given in the Ref.~\cite{atoms4040029}, in which data for these resonant channels are deduced from~\cite{PhysRevA.73.022701}.

Another very important application in nuclear fusion field is related to negative ion sources for neutral beam injection (NBI) system. In particular, electron-D$_2$ vibrationally resolved processes are paramount importance for kinetic models~\cite{Taccogna_2017} simulating the production, transport, extraction, acceleration and neutralization of negative ion beams to heat thermonuclear reactors. In this regard, a relevant result is that sources operated with deuterium could not achieve the same performances as a source operated with hydrogen, at the same power and pressure~\cite{Bacal_2020}, indicating a strong isotopic dependence from nuclear motion effects in these resonance channels, quite distinct from non-resonant electron collision systems. A recent detailed  experimental study of this well established isotope effect in dissociative attachment resonance channels is given by E.~Krishnakumar \textit{et al.}~\cite{PhysRevLett.106.243201}.

A complete and updated set of deuterium cross sections is crucial to model the plasma of negative ion sources for fusion, which are currently routinely operated both with hydrogen and deuterium~\cite{doi:10.1063/1.5133076}. Since the availability and access of plasma diagnostics in these devices is limited due to construction and operation constraints imposed by source optimization, it is difficult to precisely monitor plasma parameters in the entire source volume. A small size low pressure helicon plasma such as RAID~\cite{10.1051/epjconf/201715703014}, able to achieve high power hydrogen and deuterium discharges and mimic the electron temperature and density conditions of both the driver and the expansion region of large negative ion sources for fusion, represents a versatile testbench to validate numerical models. Dedicated Particle-in-Cell (PiC) and fluid modeling are currently underway to shed light on transport and chemistry of hydrogen RAID plasma discharges~\cite{fubiani2021}.


Aimed at supplying data for non-equilibrium plasma modeling applications~\cite{TACCOGNA201227, atoms5020018, WuenderlichFantz2016}, a large number of theoretical~\cite{Meltzer_2020, SCARLETT2021101361, PhysRevA.90.022711, Tapley_2018} and experimental~\cite{Allan_1985, Wingerden_1977, doi:10.1063/1.1696701, Wrkich_2002} works, including a review~\cite{doi:10.1063/1.555856}, have appeared for H$_2$ cross sections. Moreover, recently, results for H$_2^+$ collisions with electrons~\cite{PhysRevA.88.062709, doi:10.1063/1.5090054, romanian_astro2020}, for the process of dissociative recombination, have become available. In spite of that, very little information exists in the literature about D$_2$ excitation and dissociation and, more generally, for the nuclear fusion relevant D$_2$, DT and T$_2$ collision systems proceeding \textit{via} their anion resonances.

In order to fill the lack of data, in this paper, we will consider vibrationally-resolved resonant collisions by electron-impact, for the processes of vibrational excitation (VE), dissociative attachment (DA) and dissociative excitation (DE), reported in the Table~\ref{tab:reactions}, involving the ground electronic state -- $\textrm{X}\,^1\Sigma^+_g$ -- and the first two electronic excited states -- $\textrm{b}\,^3\Sigma^+_u$ and $\textrm{B}\,^1\Sigma^+_u$ -- of the D$_2$ molecule, which proceed through the low lying and the Rydberg states of D$_2^-$ resonances.

With regard to resonant scattering, it has been largely demonstrated that, for the collisions with electrons at low energy, it gives the dominant contribution in plasma kinetics~\cite{Chen2016,10.1002/9783527619078.ch4, 9783540674160}. The cross section calculations we will be performed in the so-called ``Local-Complex-Potential'' (LCP) approach~\cite{0034-4885-31-2-302, Domcke199197, PhysRevA.20.194, PhysRevA.73.032721} and for the first time this methodology will be applied for vibrational transitions among excited electronic states.

In this work, we are interested in resonant scatterings: the adiabatic nuclei (AN) approximation, often used to study electron molecule collision systems (see \textit{e.g.}, the molecular convergent close-coupling (MCCC) calculations in Ref.~\cite{SCARLETT2021101361} and references therein), cannot be applied in this context. While the Rydberg resonances -- \textit{via} high lying Rydberg states of the D$_2^-$ anion -- have received attention in recent literature, the hitherto poorly studied low lying resonances (\textit{via} the $\textrm{X}\,^2\Sigma^+_u$ state of D$_2^-$), are of particular relevance for the 0.5 to 5 eV electron temperatures in typical fusion detached divertor plasma scenarios. It is particularly this gap which we aim at in this paper with up to date and \textit{ab initio} theoretical calculations.

\begin{table}
\centering
\begin{tabular}{c@{\hspace{.5cm}}l}
\hline
Label & \hspace{6cm}Reaction
\\
\hline\hline
& \textit{Vibrational excitation}
\\
VE1 & $e(\epsilon) + \textrm{D}_2(\textrm{X}\,^1\Sigma^+_g; v) \to \textrm{D}_2^-(\textrm{X}\,^2\Sigma^+_u, \textrm{B}\,^2\Sigma^+_g, \textrm{C}\,^2\Sigma^+_g) \to e(\epsilon') + \textrm{D}_2(\textrm{X}\,^1\Sigma^+_g; v') $ 
\\
VE2 & $e(\epsilon) + \textrm{D}_2(\textrm{X}\,^1\Sigma^+_g; v)  \to \textrm{D}_2^-(\textrm{C}\,^2\Sigma^+_g) \to e(\epsilon') + \textrm{D}_2(\textrm{B}\,^1\Sigma^+_u; v') $
\\
VE3 & $e(\epsilon) + \textrm{D}_2(\textrm{B}\,^1\Sigma^+_u; v)  \to \textrm{D}_2^-(\textrm{C}\,^2\Sigma^+_g) \to e(\epsilon') + \textrm{D}_2(\textrm{B}\,^1\Sigma^+_u; v') $
\\
\\
& \textit{Dissociative electron attachment}
\\
DA1 & $ e(\epsilon) + \textrm{D}_2(\textrm{X}\,^1\Sigma^+_g; v) \to \textrm{D}_2^-(\textrm{X}\,^2\Sigma^+_u, \textrm{B}\,^2\Sigma^+_g) \to \textrm{D}(1s) + \textrm{D}^-(1s^2) $
\\
DA2 & $ e(\epsilon) + \textrm{D}_2(\textrm{X}\,^1\Sigma^+_g; v) \to \textrm{D}_2^-(\textrm{C}\,^2\Sigma^+_g) \to \textrm{D}(n=2) + \textrm{D}^-(1s^2) $
\\
DA3 & $ e(\epsilon) + \textrm{D}_2(\textrm{B}\,^1\Sigma^+_u; v) \to \textrm{D}_2^-(\textrm{C}\,^2\Sigma^+_g) \to \textrm{D}(n=2) + \textrm{D}^-(1s^2) $
\\
\\
& \textit{Dissociative excitation}
\\
DE1 & $e(\epsilon) + \textrm{D}_2(\textrm{X}\,^1\Sigma^+_g; v) \to \textrm{D}_2^-(\textrm{X}\,^2\Sigma^+_u, \textrm{B}\,^2\Sigma^+_g,  \textrm{C}\,^2\Sigma^+_g) \to e(\epsilon') + \textrm{D}_2(\textrm{X}\,^1\Sigma^+_g; \epsilon_c) \to e(\epsilon') +\textrm{D}(1s) + \textrm{D}(1s) $
\\
DE2 & $e(\epsilon) + \textrm{D}_2(\textrm{X}\,^1\Sigma^+_g; v)  \to \textrm{D}_2^-(\textrm{B}\,^2\Sigma^+_g, \textrm{C}\,^2\Sigma^+_g) \to e(\epsilon') + \textrm{D}_2(\textrm{b}\,^3\Sigma^+_u; \epsilon_c) \to e(\epsilon') +\textrm{D}(1s) + \textrm{D}(1s)  $
\\
DE3 & $e(\epsilon) + \textrm{D}_2(\textrm{X}\,^1\Sigma^+_g; v) \to \textrm{D}_2^-(\textrm{C}\,^2\Sigma^+_g) \to e(\epsilon') + \textrm{D}_2(\textrm{B}\,^1\Sigma^+_u; \epsilon_c) \to e(\epsilon') +\textrm{D}(2p) + \textrm{D}(1s) $
\\
DE4 & $e(\epsilon) + \textrm{D}_2(\textrm{B}\,^1\Sigma^+_u; v) \to \textrm{D}_2^-(\textrm{C}\,^2\Sigma^+_g) \to e(\epsilon') + \textrm{D}_2(\textrm{B}\,^1\Sigma^+_u; \epsilon_c) \to e(\epsilon') +\textrm{D}(2p) + \textrm{D}(1s) $
\\
DE5 & $e(\epsilon) + \textrm{D}_2(\textrm{B}\,^1\Sigma^+_u; v)  \to \textrm{D}_2^-(\textrm{C}\,^2\Sigma^+_g) \to e(\epsilon') + \textrm{D}_2(\textrm{b}\,^3\Sigma^+_u; \epsilon_c) \to e(\epsilon') +\textrm{D}(1s) + \textrm{D}(1s) $
\\
DE6 & $e(\epsilon) + \textrm{D}_2(\textrm{B}\,^1\Sigma^+_u; v) \to \textrm{D}_2^-(\textrm{C}\,^2\Sigma^+_g) \to e(\epsilon') + \textrm{D}_2(\textrm{X}\,^1\Sigma^+_g; \epsilon_c) \to e(\epsilon') +\textrm{D}(1s) + \textrm{D}(1s) $
\\
\hline
\end{tabular}
\caption{List of the elementary processes, vibrationally-resolved, considered in the text. $v$ and $v'$ represent the vibrational levels and $\epsilon_c$  the energy of continuum of the electronic states of D$_2$ molecule. \label{tab:reactions}}
\end{table}

The manuscript is organized as follows: in Section~\ref{sec:th_model} we give a brief overview on the Local Complex Potential theoretical model used in the calculations and on the molecular input data; in Section~\ref{sec:results} we discuss the results within a comparison with cross sections presented in literature where available and we show the isotopologue effect with H$_2$. Finally, conclusions are present in Section~\ref{sec:conc} of the paper.

\section{Theoretical model \label{sec:th_model}}

In this section, we briefly introduce the LCP model: the theoretical framework we used to describe the resonant scattering processes presented in the Table~\ref{tab:reactions}. In the phenomenology of resonant collisions, in a first step, the incident electron is temporarily captured by the neutral target molecule forming a unstable anionic system, the resonance, which is represented by the intermediate state in the reactions in the Table~\ref{tab:reactions}. After a characteristic lifetime -- related to the so-called `resonance width' --  the negative compound decays, leading to a large spectrum of different final states, competing with each other, including excitation or dissociation of the initial molecule.

The LCP model is an effective quantum \textit{ab initio} approach which takes into account the molecular nuclear dynamics and thus it is able to consider vibrational state resolved cross sections. We restrict ourselves to the major equations of LCP and for a comprehensive theoretical treatment of the resonant collisions by electron-impact, we refer back to the original papers~\cite{0034-4885-31-2-302, Domcke199197, PhysRevA.20.194, PhysRevA.73.032721} and references therein. Recently, the LCP model was widely used to determine low-energy dissociations by electron impact and vibrational excitations for the ground state of the molecules of NO~\cite{Laporta_2020, 10.1088/1361-6595/ab86d8}, CO~\cite{0963-0252-21-4-045005} and oxygen~\cite{0963-0252-22-2-025001, PhysRevA.91.012701, Laporta201644}, which  gave results in good agreement with experiment.

With respect to the reactions in the Table \ref{tab:reactions}, in the following, we will refer to the electronic states of D$_2$, by ${\cal S} = \{\textrm{X}\,^1\Sigma^+_g, \textrm{b}\,^3\Sigma^+_u, \textrm{B}\,^1\Sigma^+_u\} = \{\textrm{X}, \textrm{b}, \textrm{B}\}$ and to the D$_2^-$ resonances, by ${\cal R}  = \{\textrm{X}\,^2\Sigma^+_u, \textrm{B}\,^2\Sigma^+_g, \textrm{C}\,^2\Sigma^+_g\}=\{\textrm{X}^-, \textrm{B}^-, \textrm{C}^-\}$. The correspondence between symmetry states and reaction channels is reported in Table~\ref{tab:limits}.

\begin{table}
\centering
\begin{tabular}{c@{\hspace{.5cm}}c@{\hspace{.5cm}}c@{\hspace{.5cm}}l}
\hline
Channel & Limit & Energy (eV) & Symmetries\\
\hline
DE3, DE4 & $\mathrm{D}(2p) + \mathrm{D}(1s)$ & +10.17 &  $\mathrm{B}\,^1\Sigma^+_u$
\\
DA2, DA3 & $\mathrm{D}(n=2) + \mathrm{D}^-(1s^2)$ & +9.44 &  $\mathrm{C}\,^2\Sigma^+_g$
\\
DE1, DE2, DE5, DE6 & $\mathrm{D}(1s) + \mathrm{D}(1s)$ & 0.00 &  $\mathrm{X}\,^1\Sigma^+_g$, $\mathrm{b}\,^3\Sigma^+_u$
\\
DA1 & $\mathrm{D}(1s) + \mathrm{D}^-(1s^2)$ & --0.75 &  $\mathrm{X}\,^2\Sigma^+_u$, $\mathrm{B}\,^2\Sigma^+_g$
\\
\hline
\end{tabular}
\caption{Dissociating channels converging to the electronic states of D$_2$ and D$_2^-$ considered in the text. Asymptotic limit positions are given with respect to the $\mathrm{D}(1s) + \mathrm{D}(1s)$ threshold. \label{tab:limits}}
\end{table}

According to the LCP approach, the cross sections for the VE, DA and DE processes -- in the rest frame of a D$_2$ molecule initially in vibrational level $v$ of the electronic state $s$ and for an incident electron of energy $\epsilon$ -- are given by:
\begin{eqnarray}
\sigma^{\mathrm{VE}}_{s,v\to s',v'}(\epsilon) &=& \sum_{r\in {\cal R}} \frac{2S_r+1}{(2S_s+1)\,2} \frac{g_r}{g_s\,2} 
\frac{64\,\pi^5\,m^2}{\hslash^4}  \frac{k'}{k}\left| \langle 
\chi^{s'}_{v'} | \mathcal{V}^{s'}_r | \xi^r_{s,v} \rangle \right|^2\,, \hspace{1cm} s,s' \in {\cal S} \,, \label{eq:VExsec}
\\
\sigma^{\mathrm{DA}}_{s,v}(\epsilon) &=& \sum_{r\in {\cal R}} \frac{2S_r+1}{(2S_s+1)\,2} \frac{g_r}{g_s\,2} 
2\pi^2\,\frac{K}{\mu}\,\frac{m}{k}\,\lim_{R\to\infty}\left|\xi^r_{s,v}(R)\right|^2\,, \hspace{1cm} s \in {\cal S}  \,,  \label{eq:DAxsec}
\\
\sigma^{\mathrm{DE}}_{s,v}(\epsilon) &=& \sum_{r \in {\cal R}} \frac{2S_r+1}{(2S_s+1)\,2} \frac{g_r}{g_s\,2} 
\frac{64\,\pi^5\,m^2}{\hslash^4} 
\int_{\epsilon^{th}}^{\epsilon^{max}} d\epsilon_c \frac{k'}{k} \left | \langle 
\chi_c^{s'} | \mathcal{V}^{s'}_r | \xi^r_{s,v} \rangle \right |^2\,, \hspace{1cm} s,s' \in {\cal S} \,, \label{eq:DExsec}
\end{eqnarray}
where $2S_r+1$ and $2S_s+1$ account for the spin-multiplicities of the resonant anion state and of the neutral target state respectively, $g_r$ and $g_s$ represent the corresponding degeneracy factors, $m$ is the electron mass, $\mu$ is the reduced mass of D$_2$ nuclei, and $k$ represents the incoming electron momentum. For the VE and DE processes $k'$ is the outgoing electron momentum whereas for the DA process 
$K$ is the asymptotic momentum of the final dissociating fragments $\mathrm{D} + \mathrm{D}^-$. $\chi^{s'}_{v'}$ in Eq.~(\ref{eq:VExsec}) and $\chi^{s'}_c$ in Eq.~(\ref{eq:DExsec}) refer to the vibrational and continuum final wave function belonging to the electronic state $s'$ of D$_2$ respectively. $\langle \cdot  | \cdot | \cdot \rangle$ stands for integration over the internuclear coordinate $R$ and the integral in the DE cross section extends into the continuum part of the D$_2$ potential from the dissociation threshold energy $\epsilon^{th}$, corresponding 
to the initial vibrational level $v$, up to $\epsilon^{max} = \epsilon^{th} + 50$~eV.

In Eqs.~(\ref{eq:VExsec})--(\ref{eq:DExsec}), $\xi^r_{s,v}(R)$ is the resonant wave function solution of the nuclear equation with total energy $E=\epsilon^s_v+\epsilon$:
\begin{equation}
\left[ -\frac{\hslash^2}{2\mu}\frac{d^2}{dR^2} + V_r^-(R)  - 
\frac{i}{2}\Gamma_r(R) - E \right]\xi^r_{s,v}(R) = -\mathcal{V}^{s}_r(R)\,\chi^s_v(R)\,, 
\hspace{1cm} r\in {\cal R}  \,, \label{eq:res_wf}
\end{equation}
where $V_r^-(R)$ and $\Gamma_r(R)$ represent the potential energies and the autoionization widths for the D$_2^-$ resonant states included in the calculations and $\chi^s_v$ is the wave function for the initial vibrational state of electronic state $s$ of D$_2$ with energy $\epsilon^s_v$.

In the LCP formalism, $\mathcal{V}^s_r$ is the discrete-to-continuum coupling between the resonance $r$ and the electronic state $s$ of target given by:
\begin{equation}
{\mathcal{V}^s_r}^2 = \frac{\hslash}{2\pi} \frac{\Gamma^s_r}{k}\,,\hspace{1cm} r\in {\cal R} \,,  \label{eq:couplingVr}
\end{equation}
where $\Gamma^s_r$ is the partial width of the resonance $r$ with respect to the electronic state $s$ of the D$_2$ molecule. The relationship with the total width present in the Eq.~(\ref{eq:res_wf}) is given by:
\begin{equation}
\Gamma_r = \sum_{s \in {\cal S}} \Gamma_r^s \,,\hspace{1cm} r \in {\cal R} \,.
\end{equation}

In the calculations, we used the potential energy curves of H$_2$ molecule for the ground state $\textrm{X}\,^1\Sigma^+_g$ \cite{doi:10.1063/1.1725796, SHARP1970119}, for the first repulsive excited state $\textrm{b}\,^3\Sigma^+_u$ \cite{doi:10.1063/1.1697142} and for the $\textrm{B}\,^1\Sigma^+_u$ \cite{doi:10.1063/1.453888} excited state but with a reduced mass $\mu=1835.741$~a.u of D$_2$ nuclei. Concerning for the resonant states D$_2^-$, the complex potential for the lowest valence was taken from papers \cite{BardsleyWadehra1979, PhysRevA.88.062701} whereas the data for the Rydberg excited state $\mathrm{C}\,^2\Sigma^+_g$ come from the R-matrix calculation contained in the paper \cite{0953-4075-31-4-027}.

\begin{figure}
\centering
\begin{tabular}{ m{10cm} m{7cm} }
\includegraphics[scale=.51]{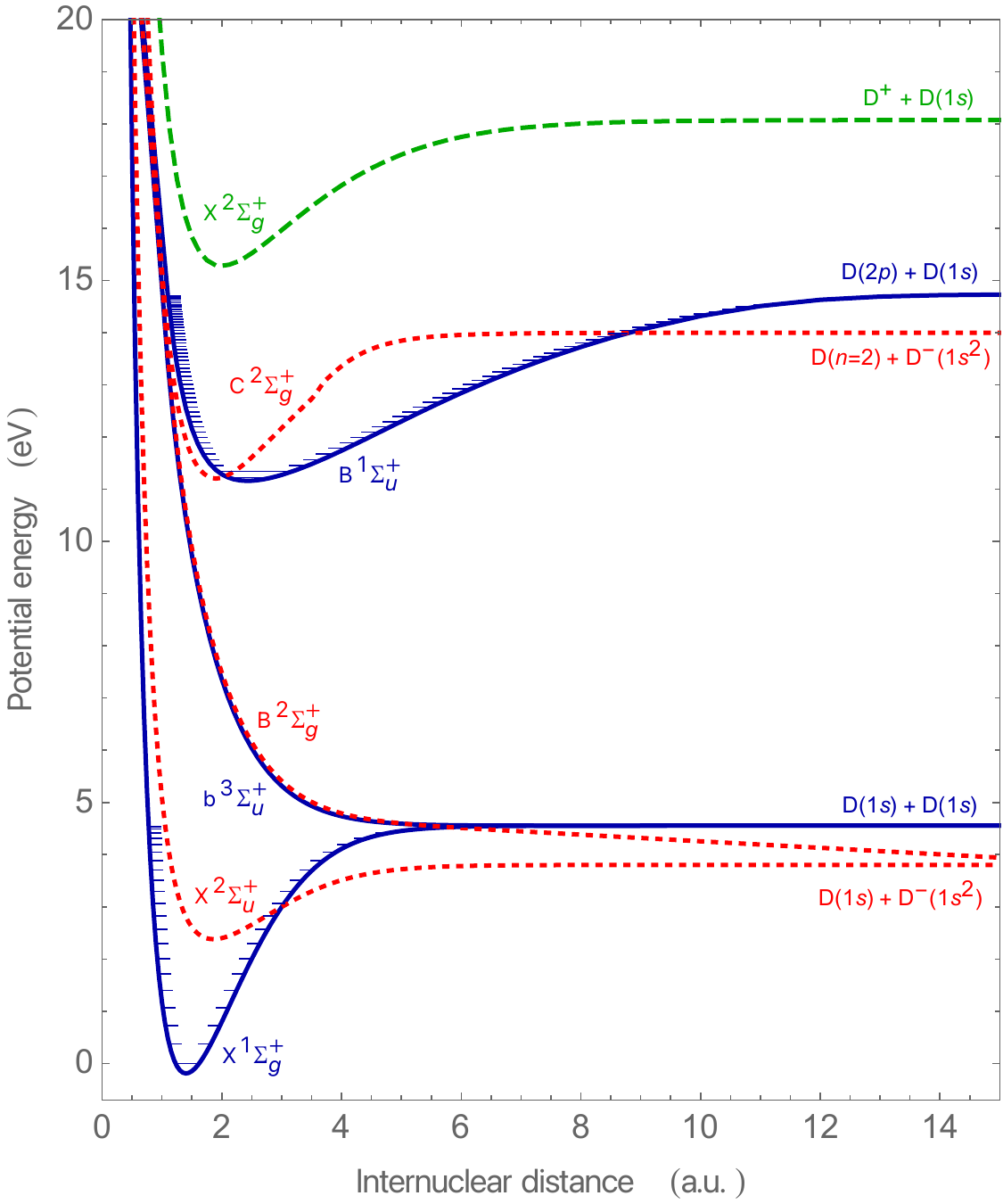}  & 
\includegraphics[scale=.3]{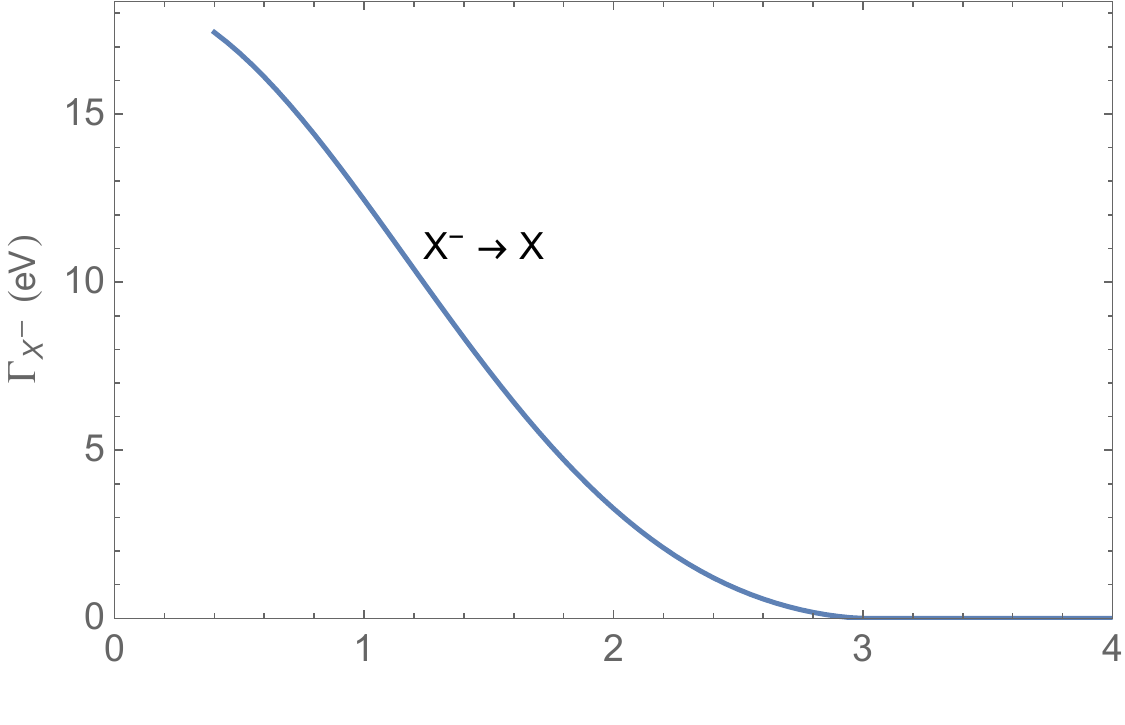} \includegraphics[scale=.3]{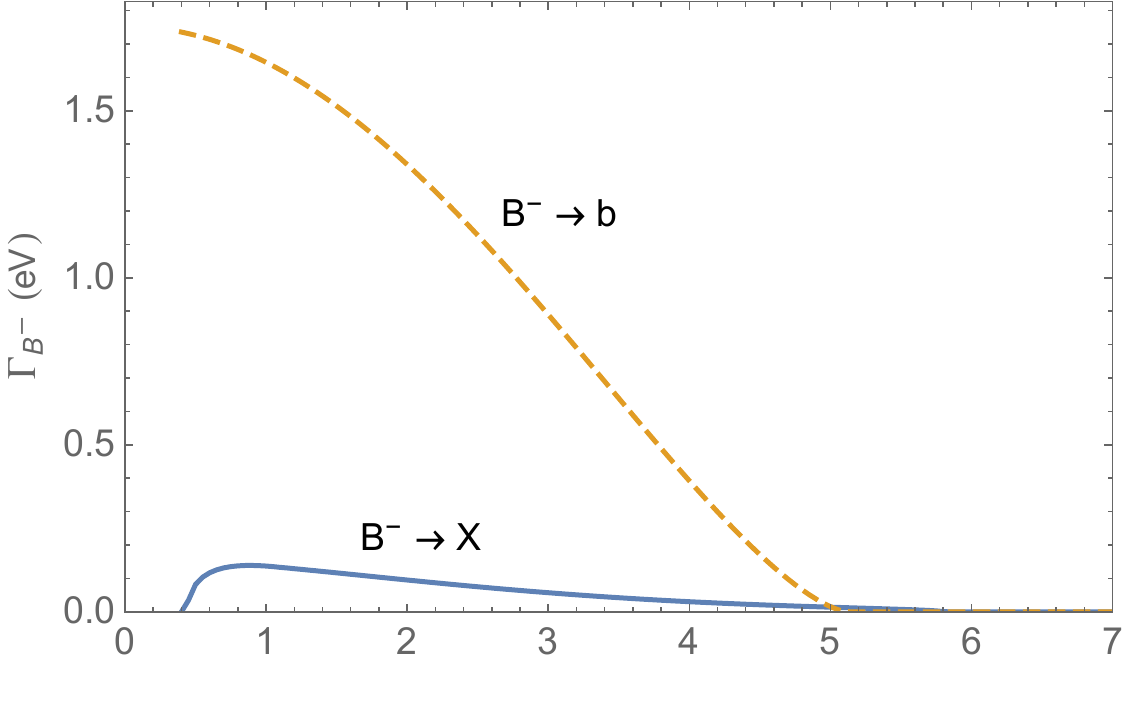}  \includegraphics[scale=.3]{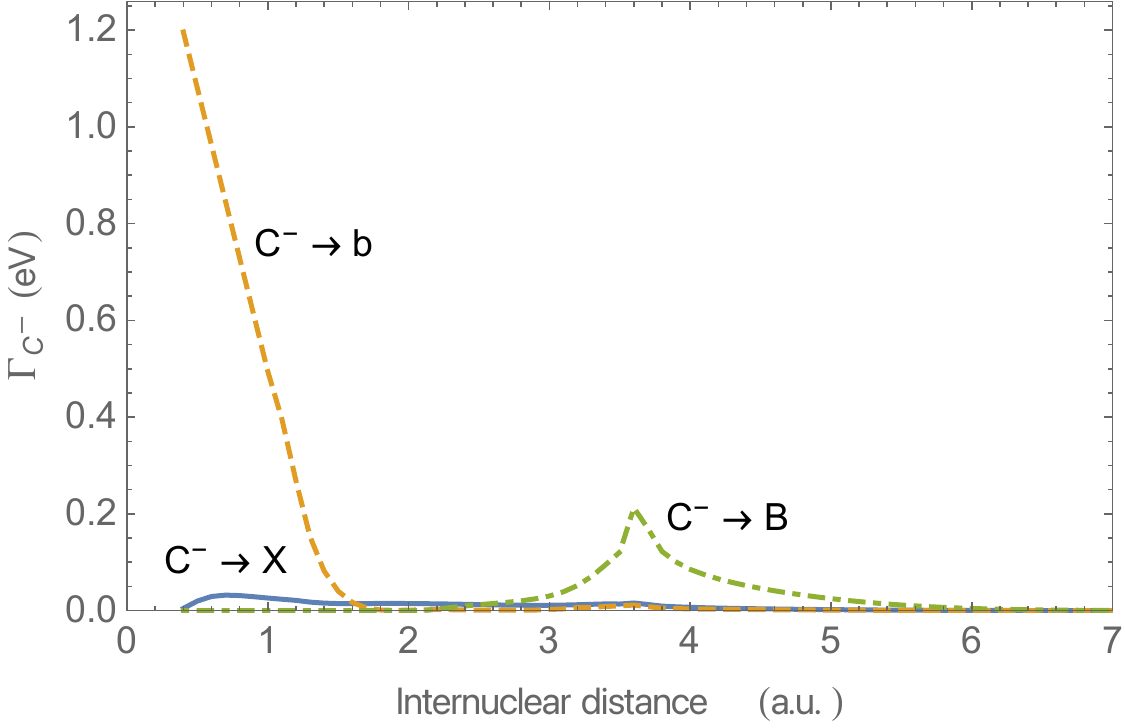} 
\end{tabular}
\caption{(Plot on the left) Potential energy curves of D$_2$ electronic states (solid blue lines) and of D$_2^-$ resonances (dotted red lines) considered in the calculations. For sake of comparison the ground state of D$_2^+$ (broken green line) is also shown. (Plots on the right) Partial widths with respect to the neutral states of the three resonances included in the calculations as a function of the internuclear distance. \label{fig:potwidth}}
\end{figure}

Figure~\ref{fig:potwidth} summarizes the molecular data -- the potentials for D$_2$ and D$^-_2$ as well as the partial widths for the three resonant states -- used in the calculations. In the theoretical model, integration over internuclear distances carrier out over the interval $R\in [0.4,15]$~a.u. The vibrational levels and the dissociation energies for the $\textrm{X}\,^1\Sigma^+_g$ and $\textrm{B}\,^1\Sigma^+_u$ states of D$_2$ molecule are reported in Table~\ref{tab:D2_viblev}.  Potentials in Figure~\ref{fig:potwidth} and vibrational levels in Table~\ref{tab:D2_viblev} refer to rotational ground states ($j=0$) in the present work, and all cross sections in the following correspond to rotational temperature $T_R=0$, \textit{i.e.} $j=j'=0$ for all transitions.

\begin{table}
\begin{tabular}{cc|cccc}
\hline
\multicolumn{2}{c}{D$_2(\mathrm{X}\,^1\Sigma^+_g)$} & \multicolumn{4}{c}{D$_2(\mathrm{B}\,^1\Sigma^+_u)$} \\
\hline
\multicolumn{6}{c}{$\mu=1835.741$ a.u.} \\
\multicolumn{2}{c}{$R_{eq}=1.401$~a.u.} & \multicolumn{4}{c}{$R_{eq}=2.417$~a.u.} \\
\multicolumn{2}{c}{$D_e=4.747$~eV} & \multicolumn{4}{c}{$D_e=3.568$~eV} \\
\multicolumn{2}{c}{$D_0=4.555$~eV} & \multicolumn{4}{c}{$D_0=3.508$~eV} \\
\hline
~~~$v$~~~ & $~~\epsilon^{\mathrm{X}}_{v}$~(eV)~~ & ~~~$v$~~~ & $~~\epsilon^{\mathrm{B}}_{v}$~(eV)~~& ~~~$v$~~~ & $~~\epsilon^{\mathrm{B}}_{v}$~(eV)~~\\
\hline\hline
 0 &   0.000   &  0 &   11.222  & 25 &  13.536  \\
 1 &   0.371   &  1 &   11.339  & 26 &  13.605  \\
 2 &   0.727   &  2 &   11.454  & 27 &  13.673  \\
 3 &   1.069   &  3 &   11.566  & 28 &  13.738  \\
 4 &   1.397   &  4 &   11.677  & 29 &  13.802  \\
 5 &   1.710   &  5 &   11.785  & 30 &  13.865  \\
 6 &   2.010   &  6 &   11.891  & 31 &  13.926  \\
 7 &   2.295   &  7 &   11.994  & 32 &  13.986  \\
 8 &   2.566   &  8 &   12.096  & 33 &  14.043  \\
 9 &   2.823   &  9 &   12.196  & 34 &  14.099  \\
10 &   3.066   & 10 &   12.294  & 35 &  14.154  \\
11 &   3.294   & 11 &   12.390  & 36 &  14.207  \\
12 &   3.506   & 12 &   12.484  & 37 &  14.258  \\
13 &   3.703   & 13 &   12.575  & 38 &  14.308  \\
14 &   3.883   & 14 &   12.666  & 39 &  14.357  \\
15 &   4.045   & 15 &   12.754  & 40 &  14.403  \\
16 &   4.189   & 16 &   12.840  & 41 &  14.448  \\
17 &   4.312   & 17 &   12.924  & 42 &  14.491  \\
18 &   4.414   & 18 &   13.007  & 43 &  14.532  \\
19 &   4.490   & 19 &   13.088  & 44 &  14.571  \\
20 &   4.538   & 20 &   13.167  & 45 &  14.606  \\
   &            & 21 &   13.244  & 46 &  14.640  \\
   &            & 22 &   13.320  & 47 &  14.670  \\
   &            & 23 &   13.393  & 48 &  14.696  \\
   &            & 24 &   13.466  & 49 &  14.718  \\
\hline
\end{tabular}
\caption{Some spectroscopic proprieties and energy of vibrational levels for the $\mathrm{X}\,^1\Sigma^+_g$ and $\mathrm{B}\,^1\Sigma^+_u$ electronic states of D$_2$ molecule.  \label{tab:D2_viblev}}
\end{table}

\section{Results and discussion \label{sec:results}}

\begin{figure}
\centering
\begin{tabular}{ccc}
& \includegraphics[scale=.4]{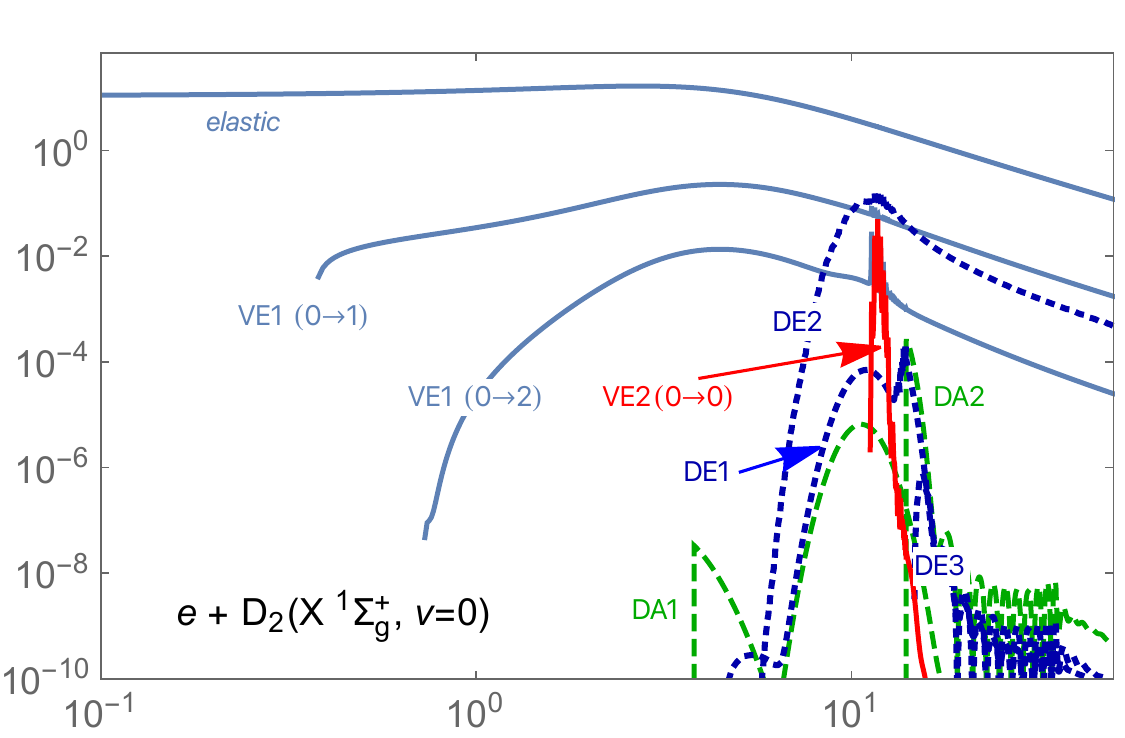} & \includegraphics[scale=.4]{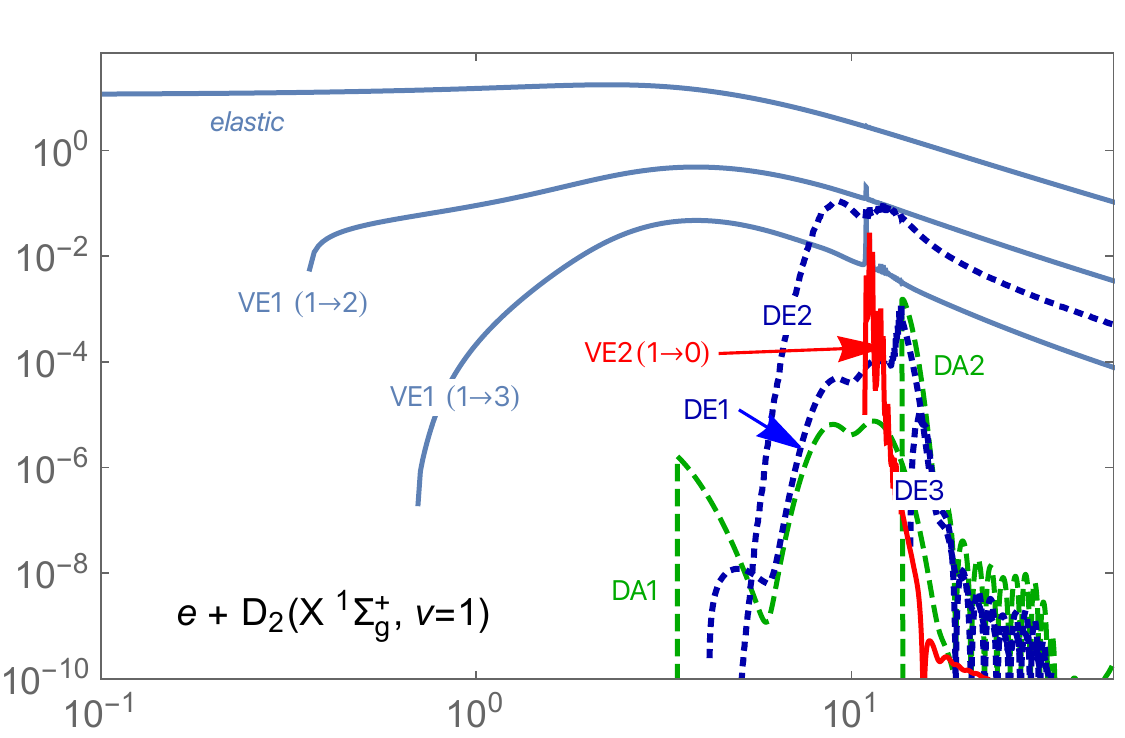} \\
\rotatebox{90}{\textsf{Cross section (\AA$^2$)}} & \includegraphics[scale=.4]{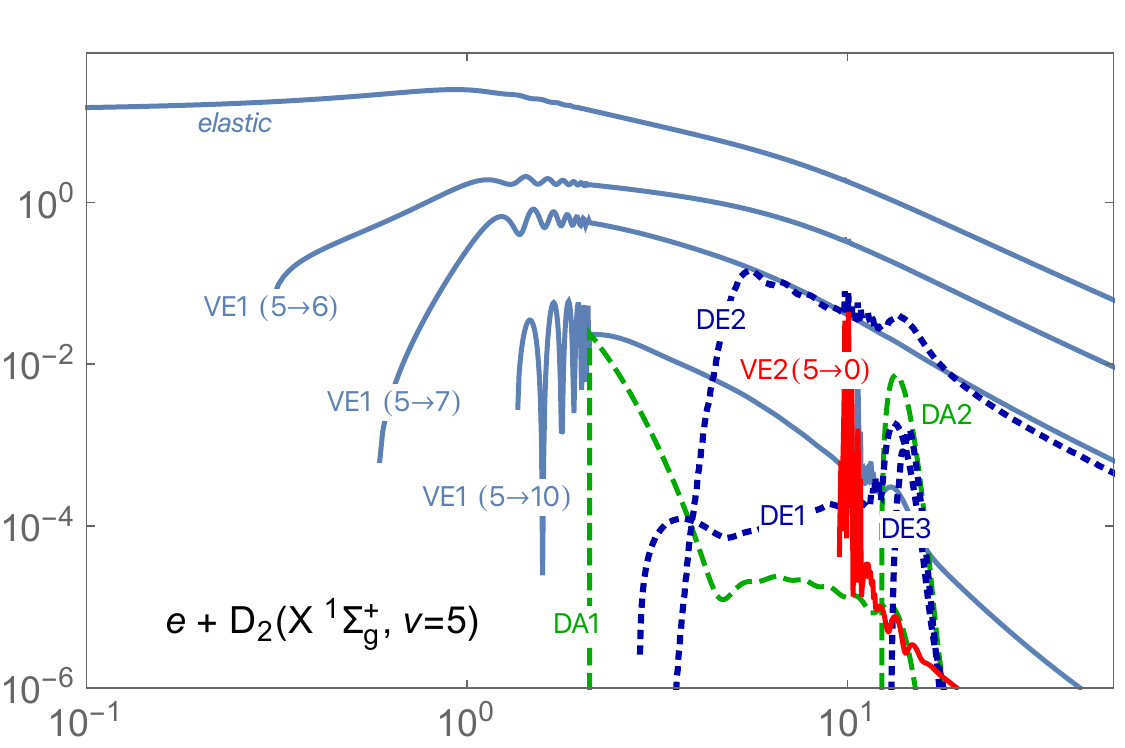} & \includegraphics[scale=.4]{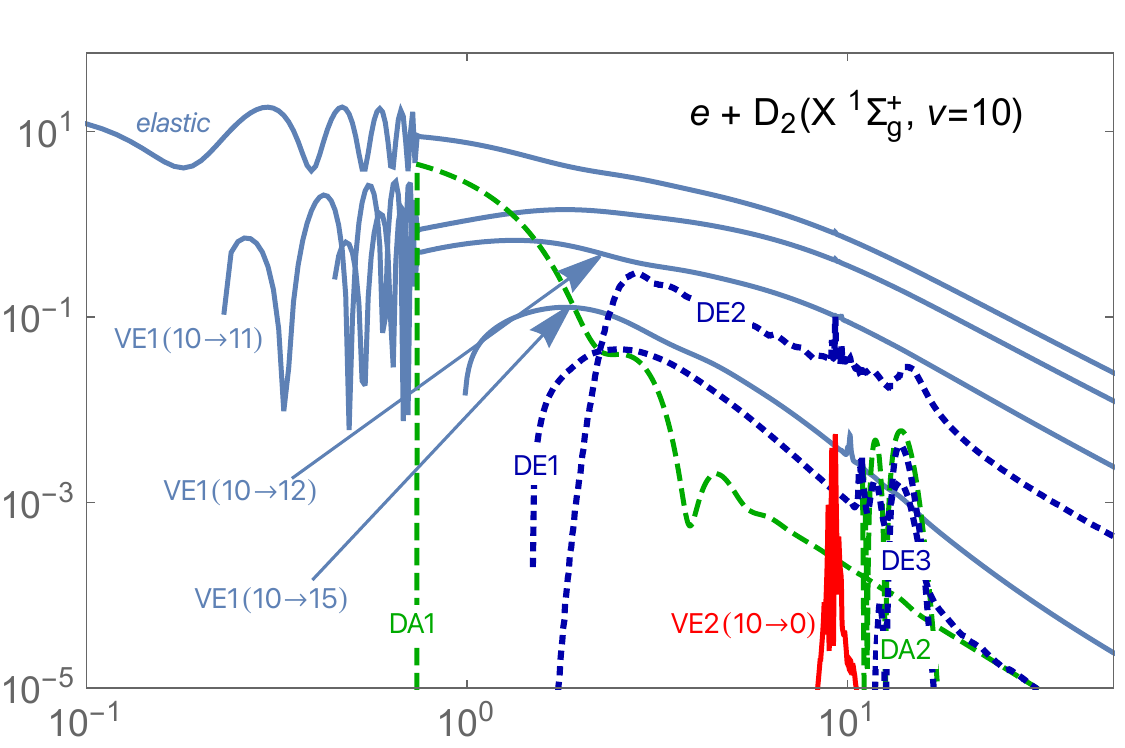} \\
& \includegraphics[scale=.4]{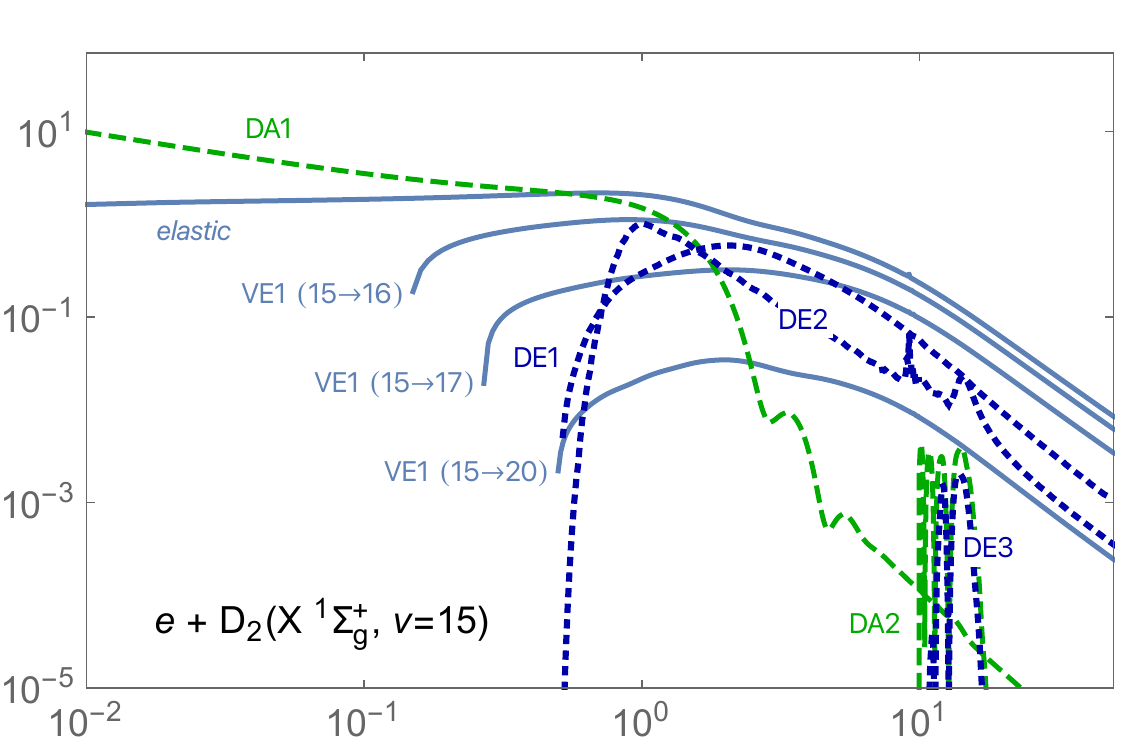} & \includegraphics[scale=.4]{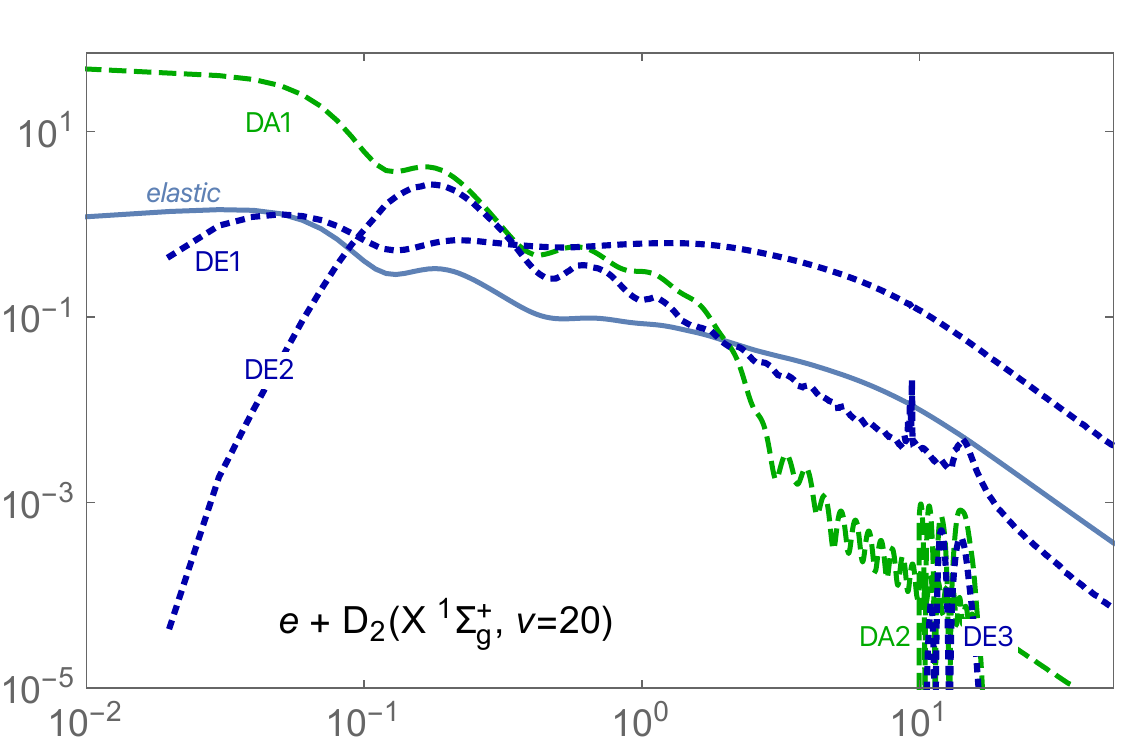} \\
& \multicolumn{2}{c}{\textsf{Incident electron energy (eV)}}
\end{tabular}
\caption{Overview of vibrationally resolved resonant cross sections by electron impact for the ground state $\textrm{X}\,^1\Sigma^+_g$ of D$_2$ molecule for the processes of vibrational excitation (elastic and VE, solid lines), dissociative attachment (DA, dashed lines) and dissociative excitation (DE, dotted lines) as obtained from the LCP model.  \label{fig:D2xsec_fullv}}
\end{figure}

\begin{figure}
\centering
\begin{tabular}{ccc}
& \includegraphics[scale=.4]{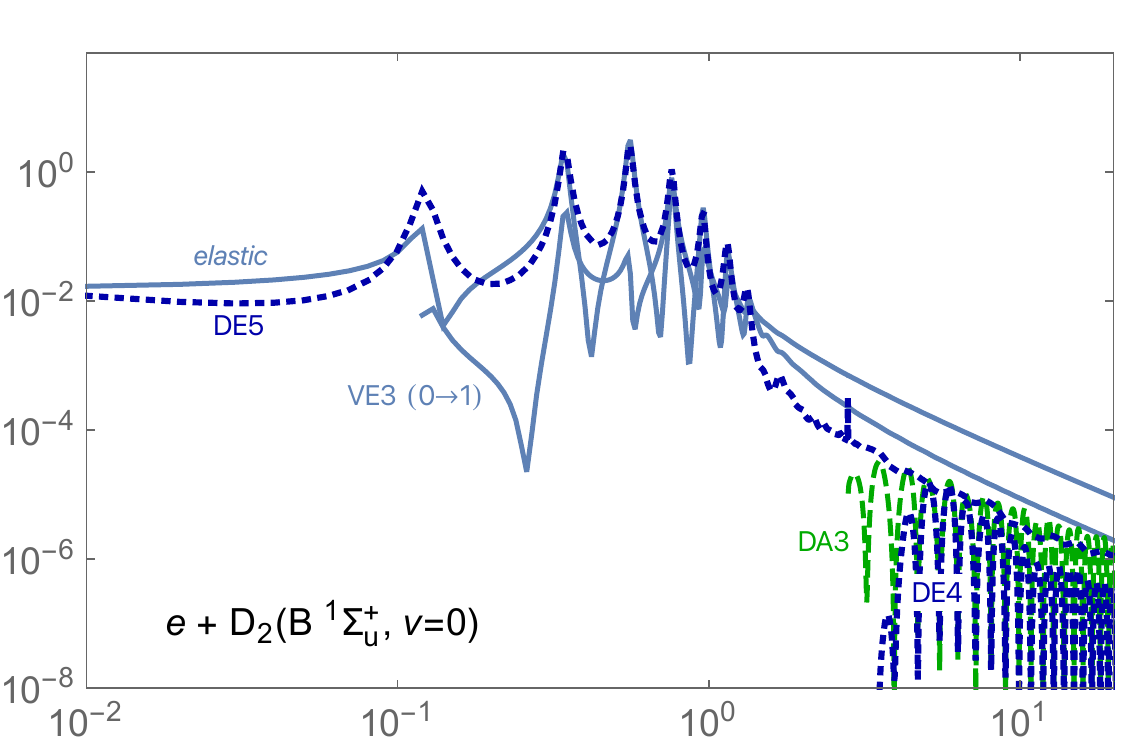} & \includegraphics[scale=.4]{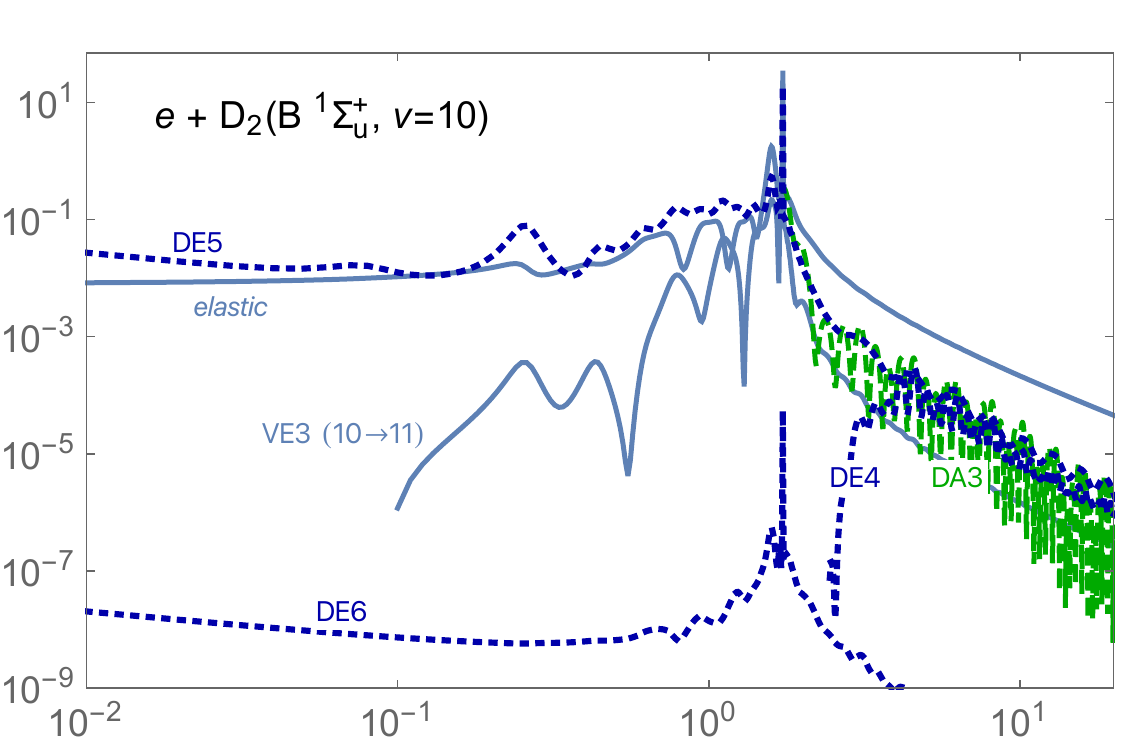} \\
\rotatebox{90}{\textsf{Cross section (\AA$^2$)}} & \includegraphics[scale=.4]{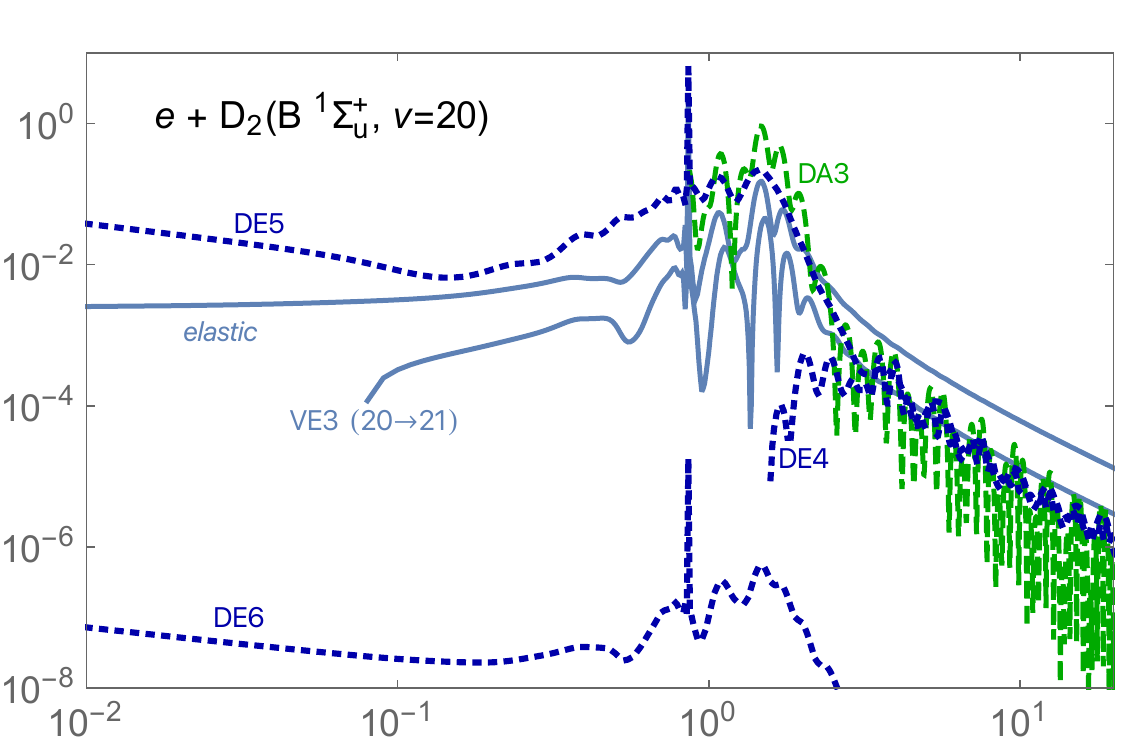} & \includegraphics[scale=.4]{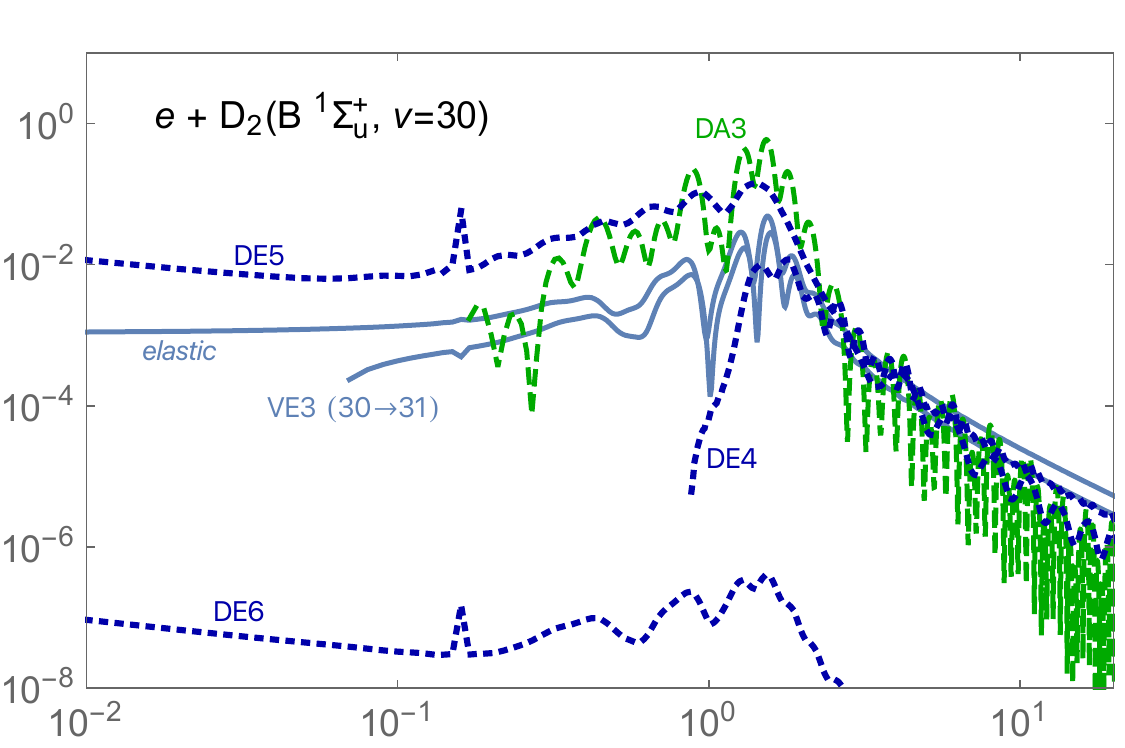} \\
& \includegraphics[scale=.4]{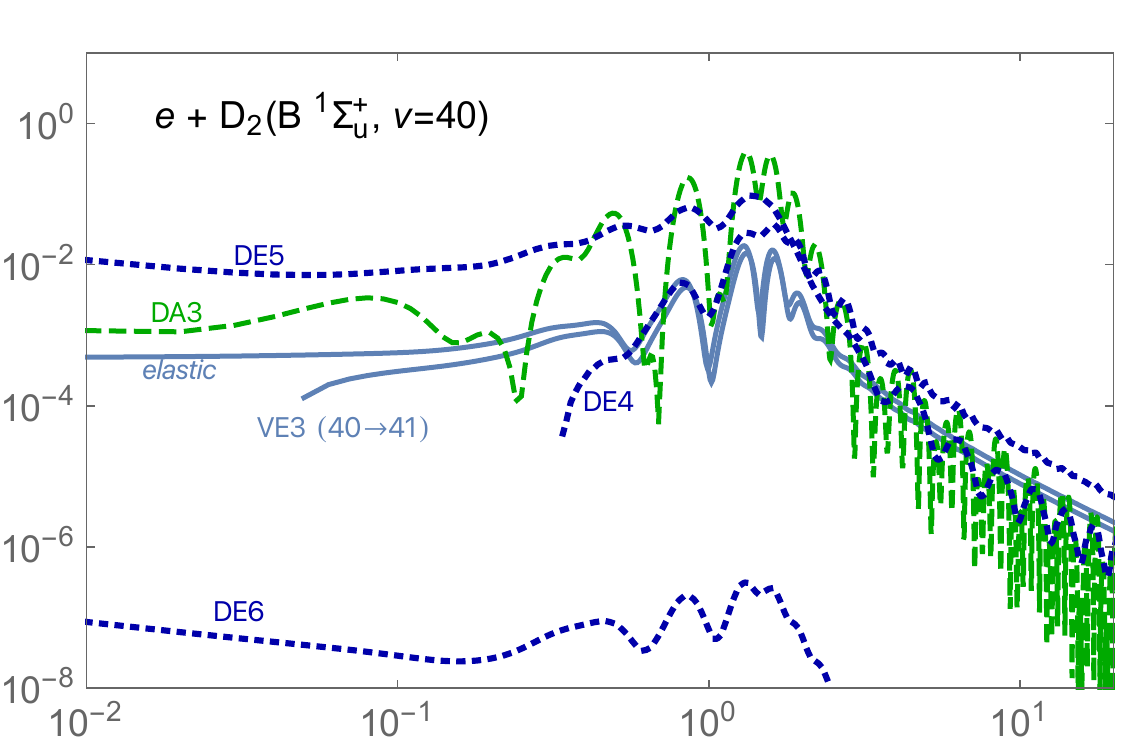} & \includegraphics[scale=.4]{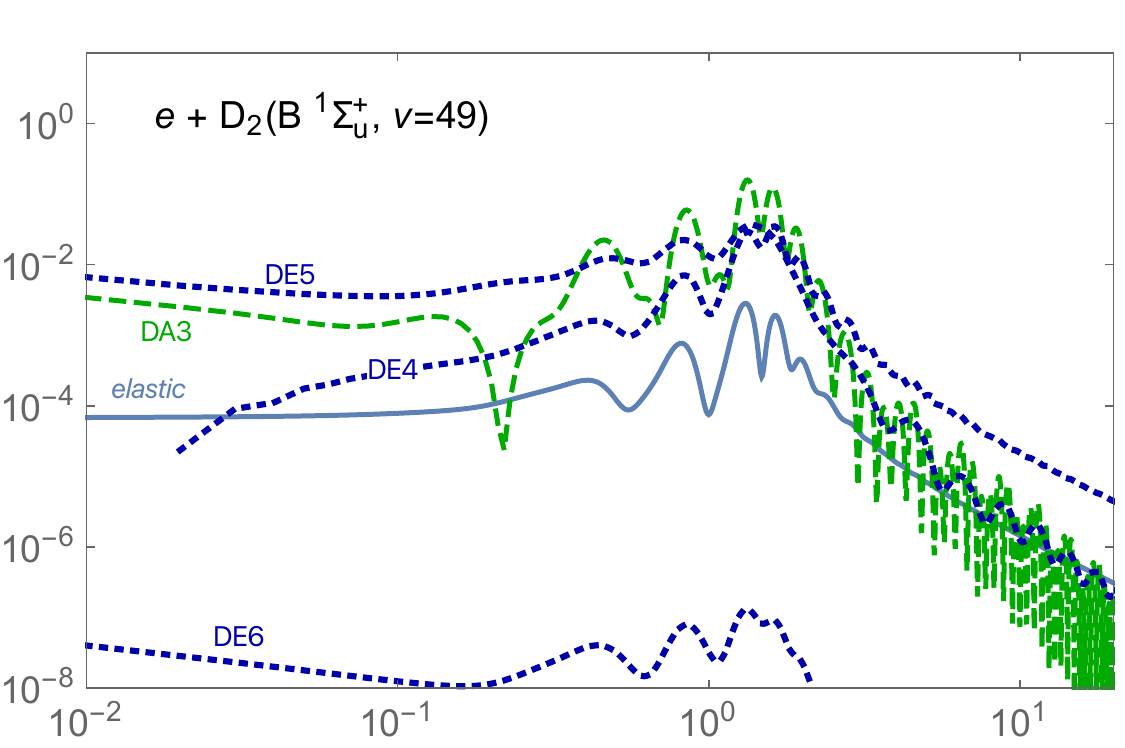} \\
& \multicolumn{2}{c}{\textsf{Incident electron energy (eV)}}
\end{tabular}
\caption{Same as in Figure~\ref{fig:D2xsec_fullv} but for the $\textrm{B}\,^1\Sigma^+_u$ excited state of D$_2$ molecule. \label{fig:D2_Bxsec_fullv}}
\end{figure}

Figure~\ref{fig:D2xsec_fullv} and Figure~\ref{fig:D2_Bxsec_fullv} summarize the results of this work. They present an overview over the full set of cross sections vibrationally resolved for the ground electronic state $\mathrm{X}\,^1\Sigma^+_g$ and for the excited electronic $\mathrm{B}\,^1\Sigma^+_u$ state of the D$_2$ molecule obtained in the LCP approach, for the processes listed in the Table~\ref{tab:reactions}. It should be noted, concerning the vibrational transitions, that only inelastic excitations, \textit{i.e.} for $v' > v$, the so-called vibrational excitations (VE), are shown in the plots: super-elastic transitions, also known as vibrational de-excitation (VdE) processes, \textit{i.e.} for $v' < v$, can be obtained from the VE by the detailed balance principle. If $\sigma^{\mathrm{VE}}_{s,v\to s',v'}(\epsilon)$ represents the cross sections for the inelastic transition, the corresponding VdE process with electron energy $\epsilon'$ is given by: 
\begin{equation}
\sigma^{\mathrm{VdE}}_{s',v' \to s,v}(\epsilon') = \sigma^{\mathrm{VE}}_{s,v\to s',v'}(\epsilon'+\epsilon^{th}_{s,v\to s',v'})\,\frac{\epsilon'+\epsilon^{th}_{s,v\to s',v'}}{\epsilon'}\,,
\end{equation}
where $\epsilon^{th}_{s,v\to s',v'} = \epsilon^{s'}_{v'} - \epsilon^{s}_v$ is the threshold for the VE process considered. 

Analogously, the associative detachment cross section, $\sigma^{\mathrm{AD}}_{s,v}$, and its inverse process of DA are linked by balance detail principle:
\begin{equation}
\sigma^{\mathrm{AD}}_{s,v} (E) = \sigma^{\mathrm{DA}}_{s,v}(E-\epsilon^s_v)\, \frac{m\,(E-\epsilon^s_v)}{\mu\,E}\,.
\end{equation}

Many features can be observed in the cross sections for the ground state presented in Figure~\ref{fig:D2xsec_fullv}. For low vibrational levels, $v<5$, the magnitudes of DA processes, DE1 and DE3 are negligible compared with mono- and bi-quantic VE transitions. The dominant dissociating channel is represented by the DE2 process, through the $\textrm{b}\,^3\Sigma^+_u$ state, because of the strong couplings $\mathrm{B}^-\to \mathrm{b}$ and $\mathrm{C}^-\to \mathrm{b}$ (see plots on the right in Figure~\ref{fig:potwidth}). As the vibrational levels increase and the first D$_2^-$ threshold is reached, all DA and DE channels become important: for $5<v<12$ they are comparable to the VE and for high vibrational levels, for $v>12$, they are the dominant processes. In particular, the DA1 channel become threshold-less processes and the dissociation from DE1 process has the same order of magnitude as DE2 channel.

Concerning specifically the vibrational excitation cross sections, many oscillating structures can be observed in the Figure \ref{fig:D2xsec_fullv}. In particular, two sets of peaks can be distinguished: the first one ends at the DA1 threshold and the second one, at higher energies, corresponds to the DA2 process. This behaviour can be traced back to the vibrational structures of the resonances D$_2^-$. On the other hand, the processes VE2, \textit{i.e.} the excitation to the $\textrm{B}\,^1\Sigma^+_u$ state from the ground state, due to the Frank-Condon factors, are inefficient for all vibrational levels and the corresponding cross sections are represented by narrow spikes around 11 eV.

Regarding the processes starting from the $\textrm{B}\,^1\Sigma^+_u$ state of D$_2$, Figure~\ref{fig:D2_Bxsec_fullv}, a remarkable feature is given by the dissociation toward the $\textrm{b}\,^3\Sigma^+_u$ state -- channel DE5 -- which is the overall dominant process: this behaviour is favored by kinetics and in particular by the strong coupling $\mathrm{C}^-\to \mathrm{b}$. On the opposite site, DE6 dissociation, which leads to the same threshold as DE5, is suppressed because of small coupling $\mathrm{C}^-\to \mathrm{X}$.

For high vibrational levels, as the threshold of the $\mathrm{C}\,^2\Sigma^+_g$ resonance is reached, the dissociative attachment process DE3 becomes important. As a matter of the fact no direct couplings exist between the $\textrm{B}\,^1\Sigma^+_u$ and the lower resonances (see plots in Figure~\ref{fig:potwidth}), dissociative attachment toward the limit $\mathrm{D}(1s) + \mathrm{D}^-(1s^2)$ is forbidden.

\begin{figure}
\centering
\includegraphics[scale=.55]{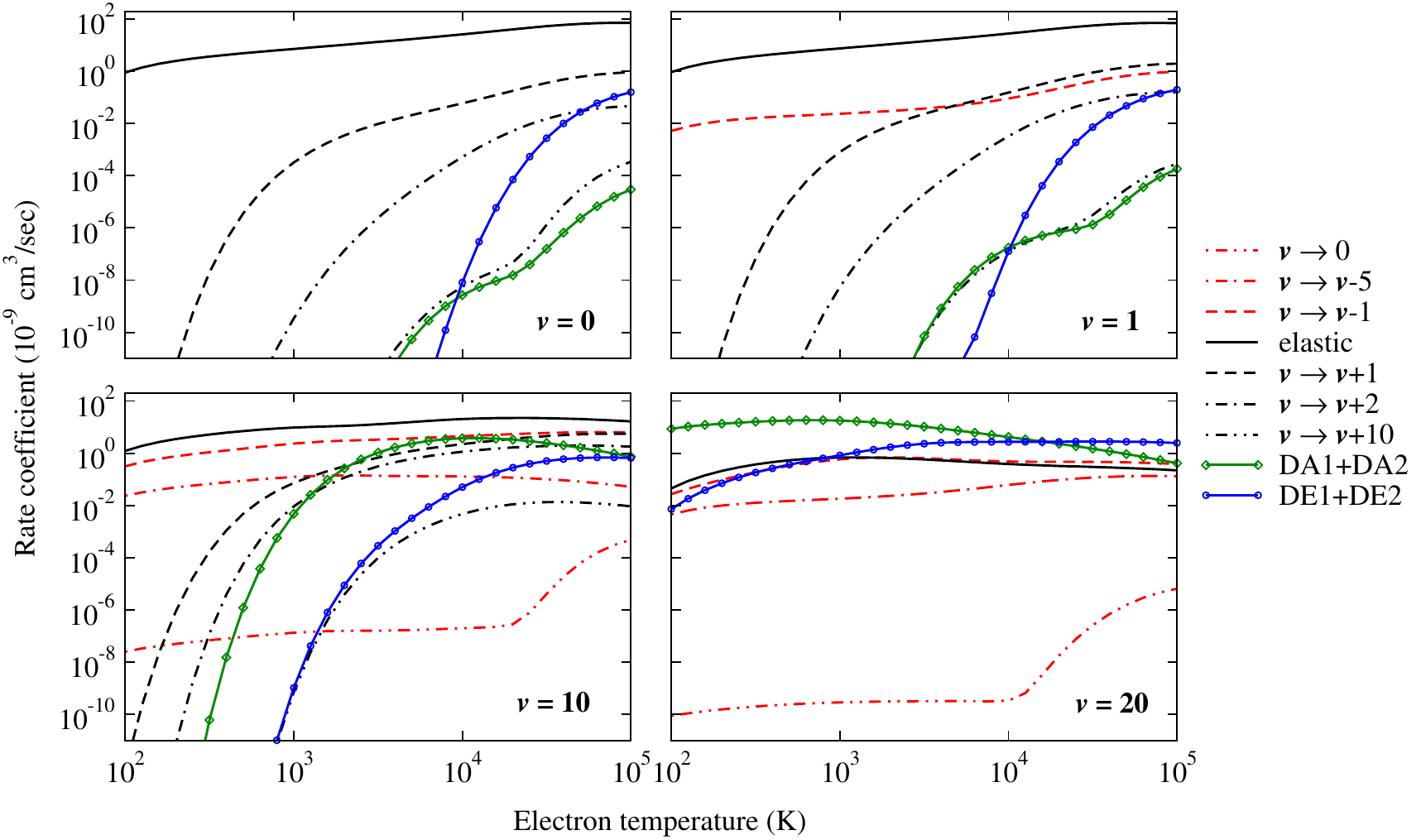} 
\caption{Selected vibrationally resolved Maxwellian rate coefficients for initial vibrational states $v=0,1,10$ and 20,  as a function of the electron temperature, for the ground state $\textrm{X}\,^1\Sigma^+_g$ of D$_2$ molecule and for the processes of vibrational excitation, dissociative attachment and dissociative excitation as obtained from the LCP model. \label{fig:D2_allrate}}
\end{figure}

To further illustrate the results of this work, Figure~\ref{fig:D2_allrate} contains some selected reaction rates (Maxwellian electron distribution), for the ground state $\textrm{X}\,^1\Sigma^+_g$ of D$_2$ molecule and for the processes reported in Table~\ref{tab:reactions}. These rates are given here exemplarily for initial vibrational states $v=0,1,10$ and 20, and plotted as a function of the electron temperature.

Globally, the behaviour of the reaction rates follows those of the corresponding cross sections. For low vibrational levels, (see \textit{e.g.} $v=0,1$), at low electron temperatures ($\lesssim 10^4$~K), multi-quantum vibrational excitations and de-excitations (broken curves) will play a important role in the plasma bulk whereas, at high temperatures ($> 10^4$~K), dissociation (DE1+DE2, curves with circles in the figure)  dominates and only mono- or bi-quantic transitions will contribute to the kinetics.

Regarding medium vibrational levels, (see \textit{e.g.} $v=10$), dissociative electron attachment process (DA1+DA2, curves with diamonds in the figure) overcomes the dissociation and for all range of electron temperatures is the major channel to dissociate the D$_2$ molecule. Vibrational transitions are important only at very low electron temperatures.

Finally, for very high vibrational levels, as expected, dissociation reactions, DA1+DA2 and DE1+DE2, dominate overall. 

\subsection{Comparison with data in the literature}

\begin{figure}
\centering
\begin{tabular}{cccc}
\multirow[c]{2}{*}[30pt]{\rotatebox{90}{\textsf{Cross section (\AA$^2$)}}}&
\includegraphics[scale=.285]{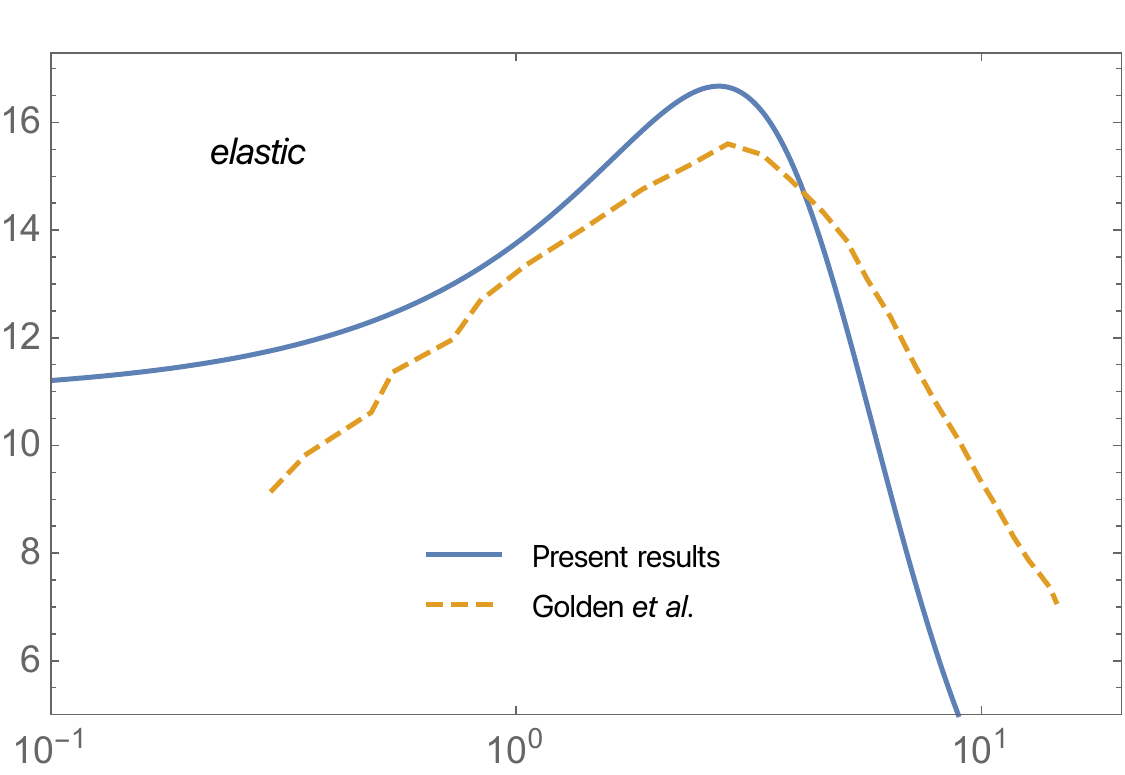} & \includegraphics[scale=.3]{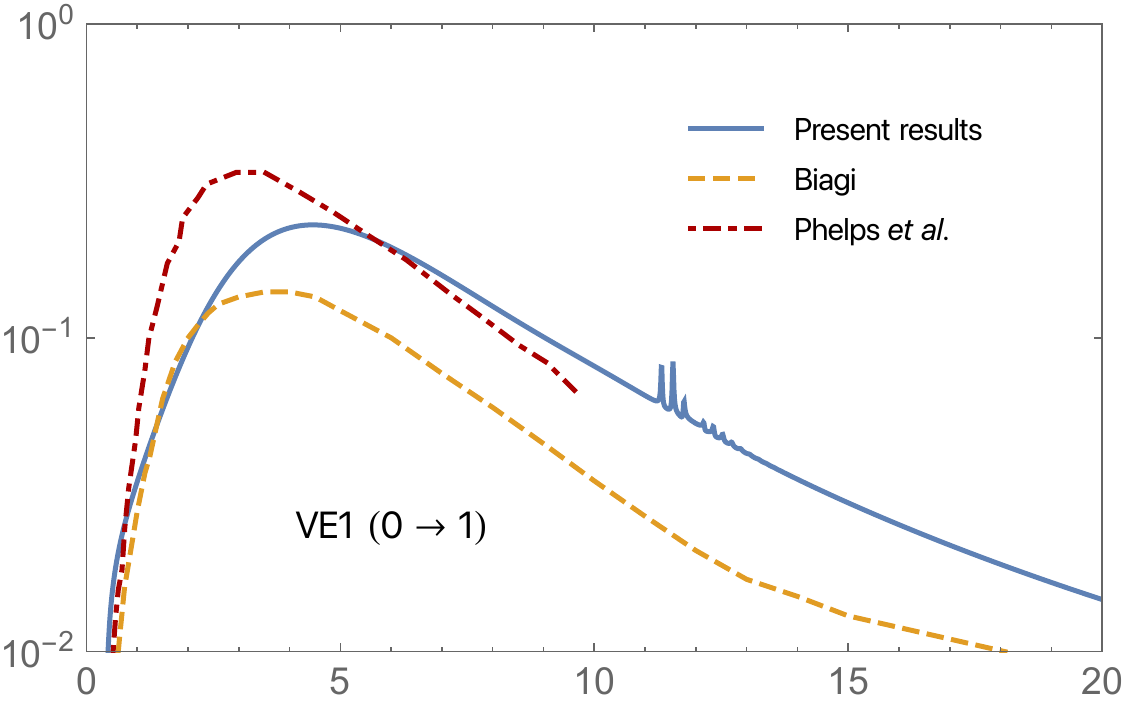} &  \includegraphics[scale=.3]{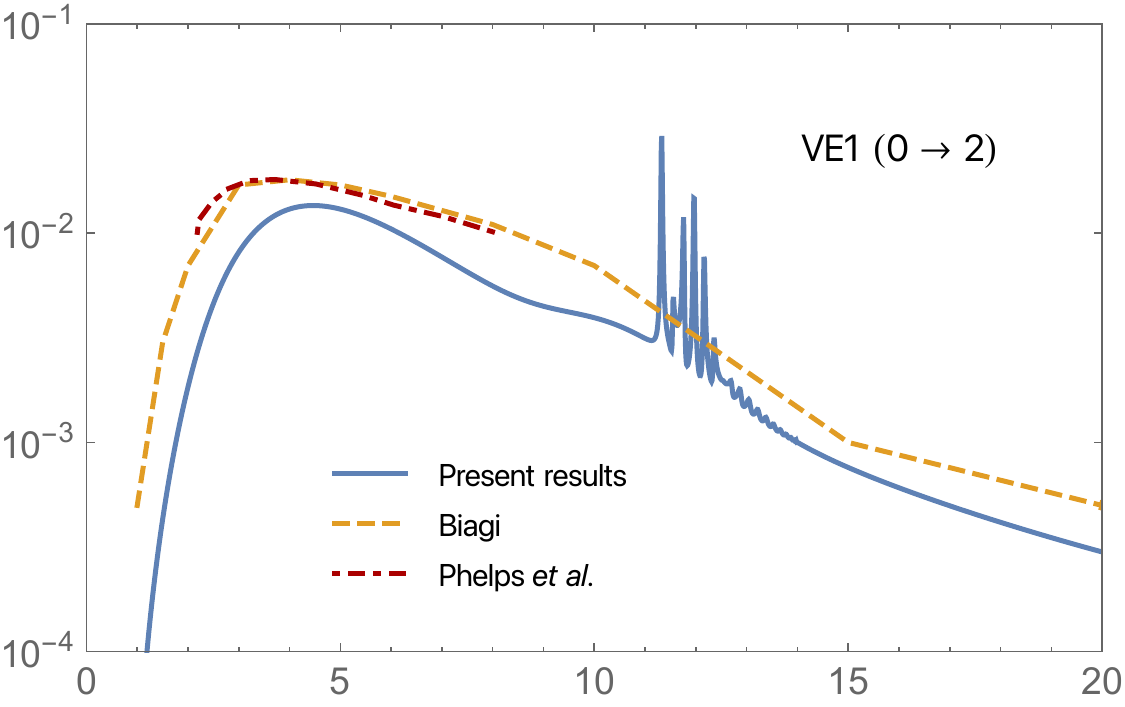}   
\\
&\includegraphics[scale=.3]{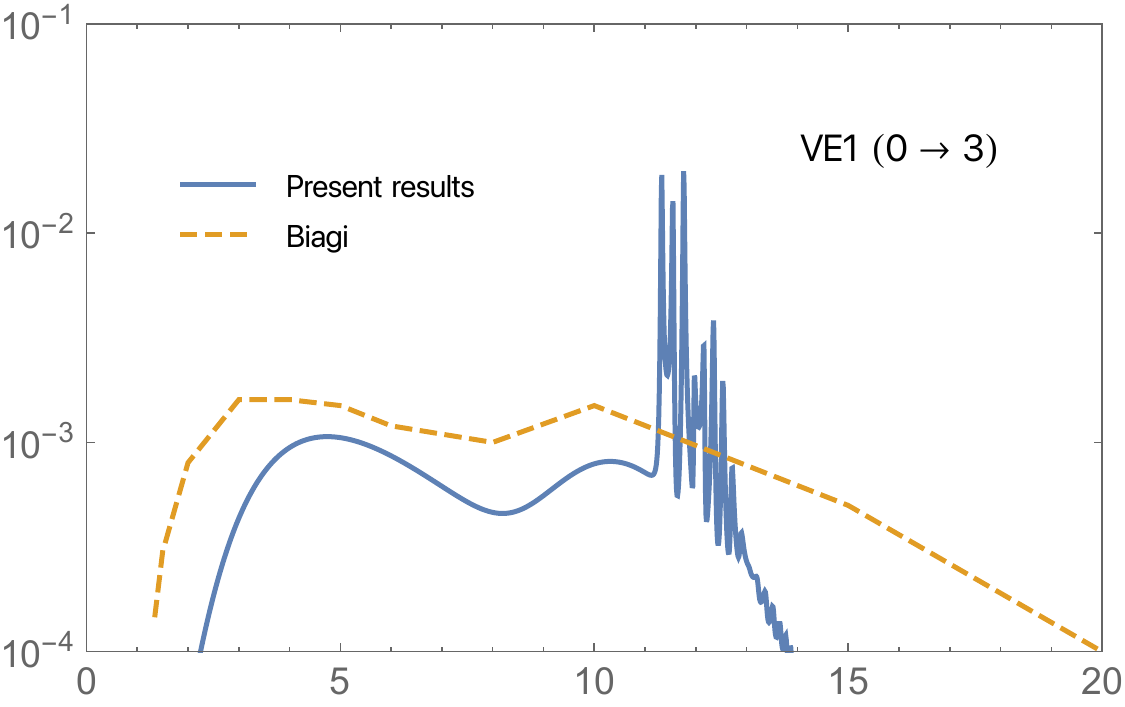} & \includegraphics[scale=.3]{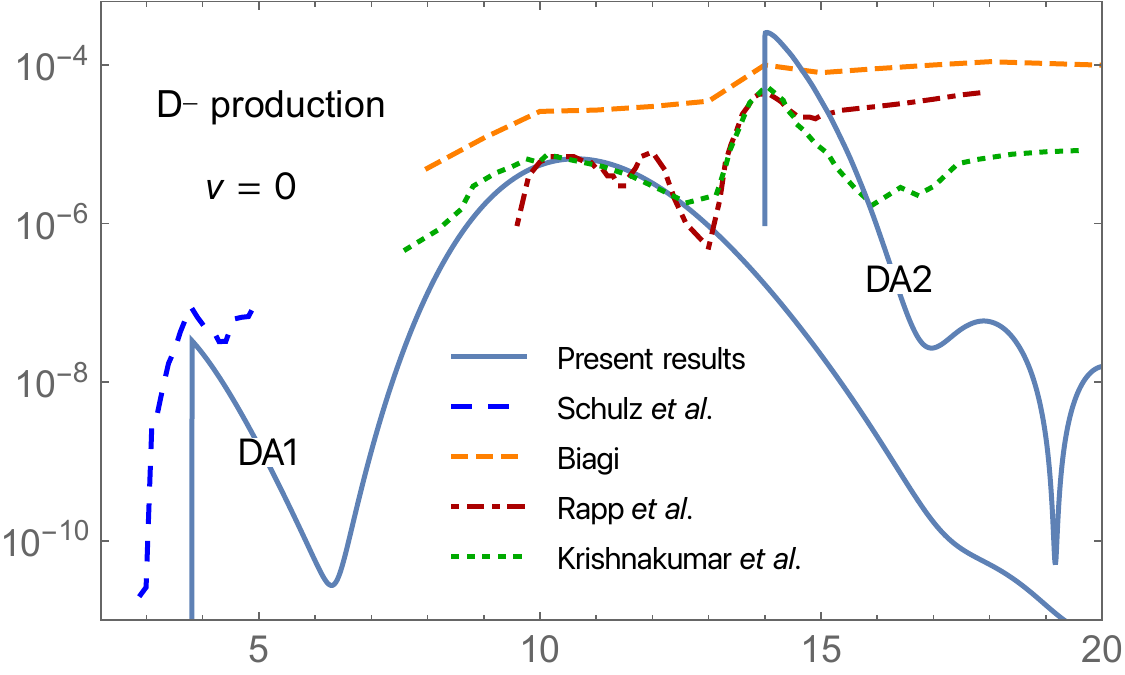} & \includegraphics[scale=.3]{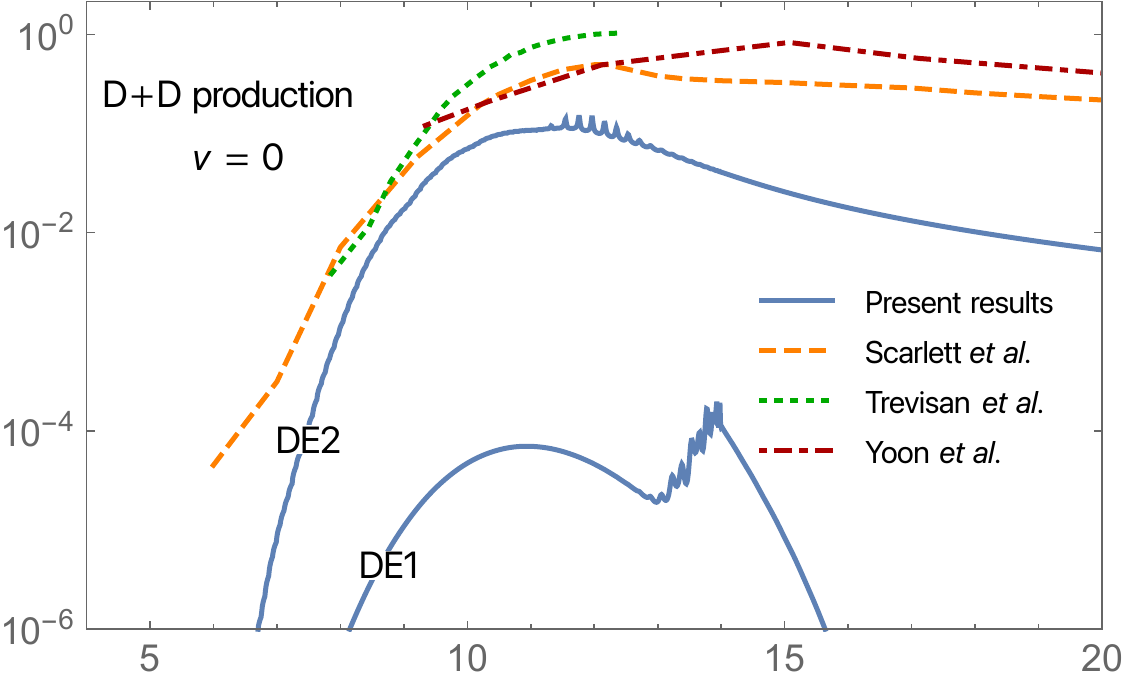} 
\\
& \multicolumn{3}{c}{\textsf{Incident electron energy (eV)}}
\end{tabular}
\caption{Comparison of cross sections obtained in the LCP approach with data present in the literature by Golden \textit{et al.}~\cite{PhysRev.146.40}, Phelps \textit{et al.}~\cite{doi:10.1063/1.448673}, Biagi \cite{doi:10.1002/ppap.201600098}, Schulz \textit{et al.}~\cite{PhysRev.158.25}, Rapp \textit{et al.}~\cite{PhysRevLett.14.533}, E.~Krishnakumar \textit{et al.}~\cite{PhysRevLett.106.243201}, Scarlett \textit{et al.}~\cite{PhysRevA.96.062708}, Trevisan \textit{et al.}~\cite{Trevisan_2002} and Yoon \textit{et al.}~\cite{Yoon_2010} for the processes indicated in the plots. \label{fig:xsec_comp}}
\end{figure}

In order to validate our LCP cross sections presented in Section~\ref{sec:results}, in Figure~\ref{fig:xsec_comp} we report comparisons with data available in the literature. In the first plot, good agreement we observed for the elastic process compared with the experimental results of Golden \textit{et al.}~\cite{PhysRev.146.40}, in particular we correctly reproduce the main peak in the cross sections which has clearly a resonant nature. As expected, LCP calculations underestimate the behaviour at high energy where non-resonant contributions become important. On the other hand, at very low energy, because nuclear spin effects are not included in our model, the LCP cross section is not able to reproduce the  anomalous quasi-elastic electron scattering~\cite{PhysRevLett.100.043204}.

Vibrational excitation cross sections -- \textit{i.e.} the plots VE1($0\to 1$), VE1($0\to 2$) and VE1($0\to 3$) -- compare well with the experiments of Buckman and Phelps \cite{doi:10.1063/1.448673} and they present a similar global shape as Biagi's calculations \cite{doi:10.1002/ppap.201600098} except for the  peaks around 12 eV due to the $\textrm{C}\,^2\Sigma^+_g$  resonance. In case of Biagi's theoretical results, unfortunately no details regarding the LXCat database nor on the theoretical method are given. The most widely used cross sections so far seem to be the ones given in Ref.~\cite{BardsleyWadehra1979}, with numerical values at 6 eV and 8 eV collision energies.

Concerning the DA processes, in the plot `D$^-$ production', we reproduce correctly the experimental results of Schulz \textit{et al.}~\cite{PhysRev.158.25} at low energies by DA1 cross section. At higher energies, our results are in agreement with experiments of Rapp \textit{et al.}~\cite{PhysRevLett.14.533}, to which we removed the background from the cross section, and with the results of E.~Krishnakumar \textit{et al.}~\cite{PhysRevLett.106.243201}. In particular we reproduce correctly the structure around 10 eV from DA1 channel and the peak at 14 eV of DA2 process.

The last plot `D+D production' compares our DE1 and DE2 cross sections with the results of Scarlett \textit{et al.}~\cite{PhysRevA.96.062708}, Trevisan \textit{et al.}~\cite{Trevisan_2002} and Yoon \textit{et al.}~\cite{Yoon_2010} for dissociative excitation process which share the same theoretical approach of `Adiabatic Nuclei' approximation and so they are able to determine non-resonant contributions. Our results reproduce, at low energies, with the dissociation of $\mathrm{b}\,^3\Sigma^+_u$ state and as expected, non-resonant scattering dominates at high energies.

In the comparisons with experimental data, we checked the discrepancy with the present results is around $15\%$. So, we can assume for the presented cross sections an uncertainty of the same order of magnitude.

\subsection{Isotopologue effect}

\begin{figure}
\centering
\includegraphics[scale=.3]{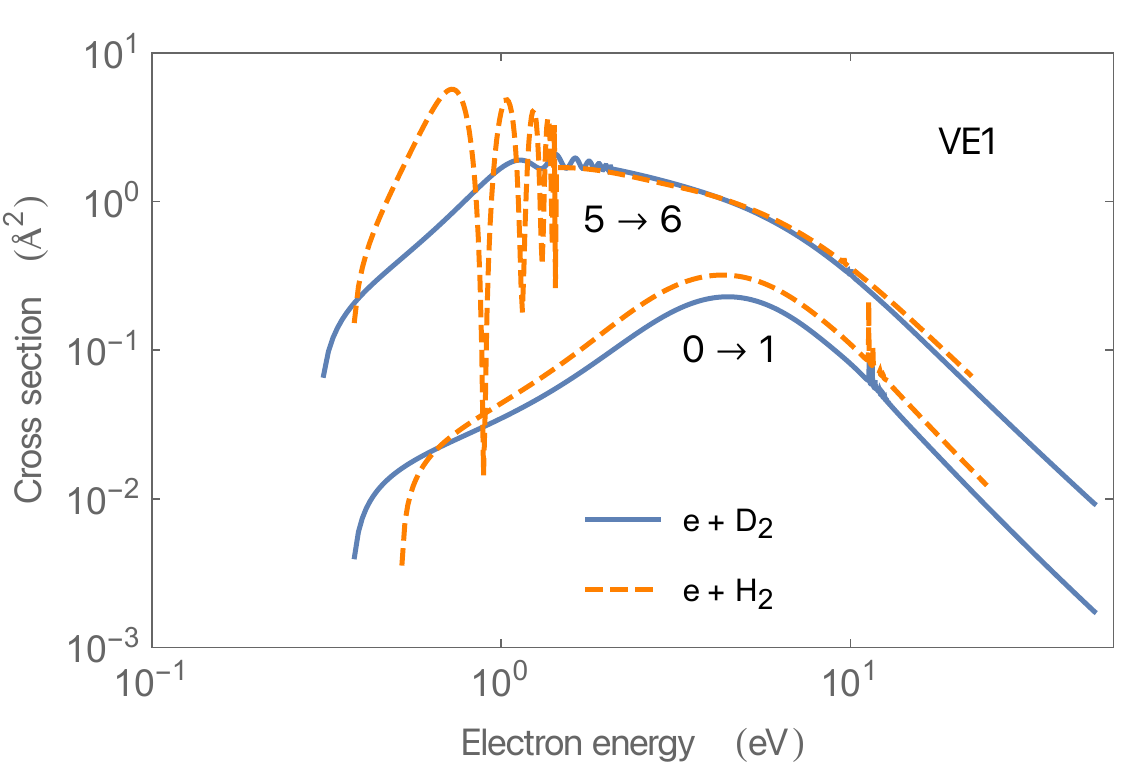}   \includegraphics[scale=.3]{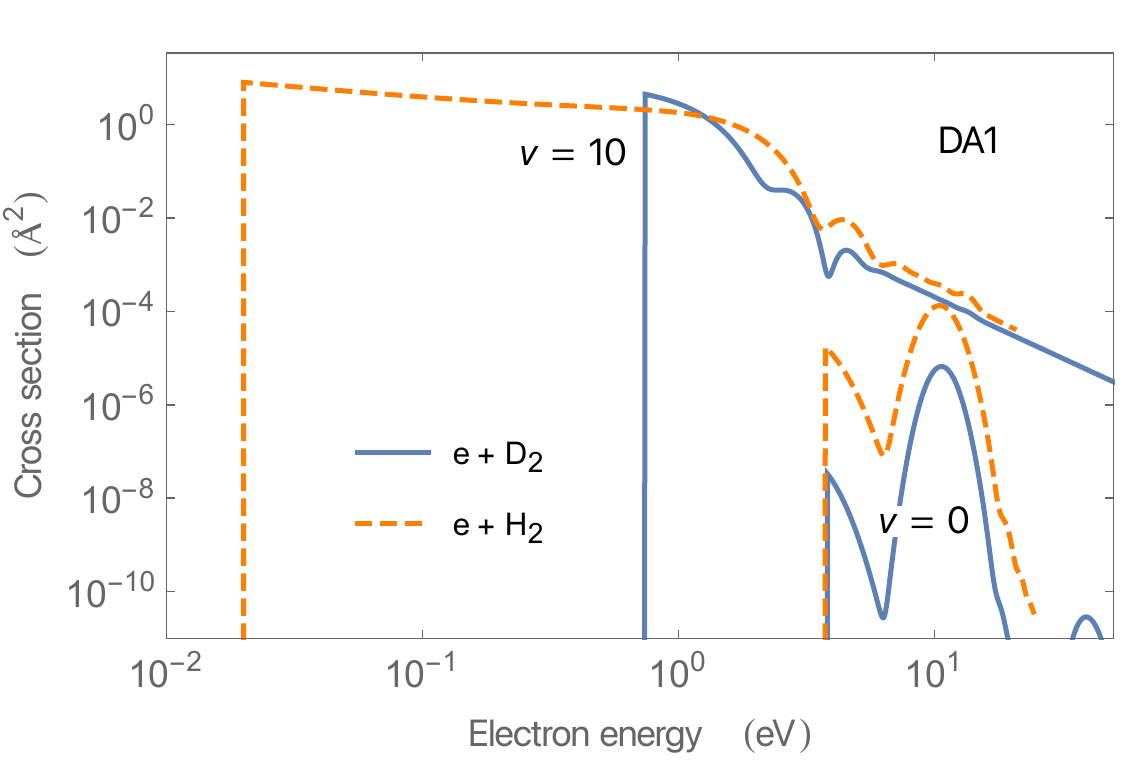}   \includegraphics[scale=.3]{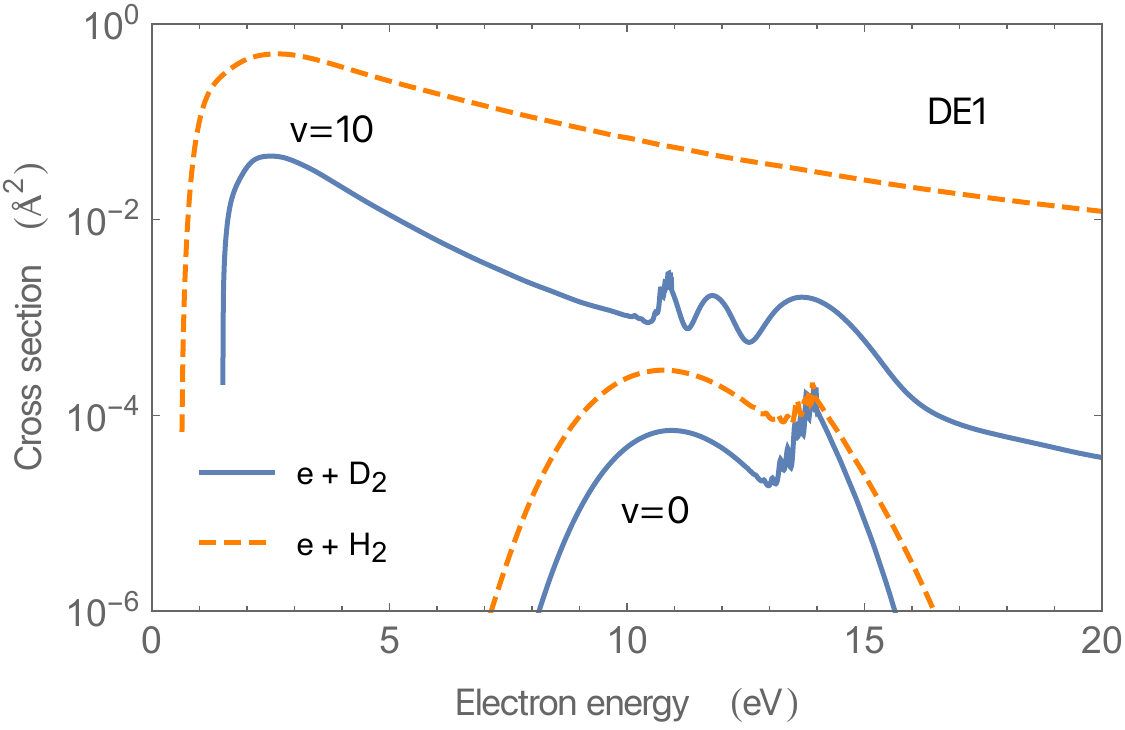}   
\caption{Isotopologue effect on cross sections of D$_2$ (solid line) and H$_2$ (dashed line) molecules for some processes of vibrational excitation, dissociative attachment and dissociative excitation. \label{fig:xsec_isot}}
\end{figure}

Plots in Figure~\ref{fig:xsec_isot} report the isotopologue effect of D$_2$ and H$_2$ molecules on cross sections for some selected processes.  We have calculated, in the same LCP theoretical approach as for D$_2$, some of the corresponding cross sections also for the H$_2$ system. A full comparative study with all cross sections will be subject to future work.

The first difference we remark is in the threshold: In fact, because the vibrational levels for H$_2$ are systematically shifted up compared to those of D$_2$ with same vibrational quantum numbers, the thresholds for DA and DE processes for H$_2$ decrease compared with those for D$_2$. On the contrary, as a consequence of the fact that the spacing between vibrational levels is smaller for D$_2$ than for H$_2$, the threshold for the VE cross sections increases for H$_2$.

Concerning the magnitude and the shape of the cross sections, in general, H$_2$ has the same trend as D$_2$ but shifted towards higher energies. However, in some cases, see for example the 5$\to$6 VE transitions, the structure of the resonant peaks is completely different. 

We should point out, in particular, the factor of about 200 difference between H$_2$ and D$_2$, in the DA cross sections for $v=0$, and for the 4 eV resonance. As this is to be regarded as the major isotope effect for fusion plasmas: no DA process in deuterium plasmas are relevant at these collision energies, but in hydrogen plasmas still some noticeable H$^-$ production (and subsequent mutual neutralization with protons) might be a relevant process channel. In fusion (divertor) plasmas one expects mostly $v=0$ (or small amounts of $v=1,2$) vibrational states. Furthermore only the lowest 4 eV resonance (first peak in DA1 cross sections) should play a role, because molecules are rapidely destroyed. This strong suppression of the lowest DA1 channel in D$_2$ as compared to H$_2$ confirms, once again, the older findings, \textit{e.g.}, already of Rapp \textit{et al.}~\cite{PhysRevLett.14.533}.

\section{Conclusions \label{sec:conc}}

In conclusion, in this work, we presented a theoretical study on vibrationally-specific cross sections for electron deuterium resonant collisions within LCP formalism. We took into account three states of neutral deuterium molecule -- including the ground $\mathrm{X}\,^1\Sigma^+_g$, the dissociative $\textrm{b}\,^3\Sigma^+_u$ and the electronic $\textrm{B}\,^1\Sigma^+_u$ states -- and three resonances of D$_2^-$ anion. In our analysis, we considered the elementary processes listed in the Table~\ref{tab:reactions}. In particular, for the first time, the LCP approach has been applied to study vibrational transitions and dissociation processes between different electronic states.

We observed the magnitude of the cross sections and the relative importance of the reactions varying over a huge range of energy and in particular as a function of the initial electronic state and vibrational levels. This behaviour relies basically on the specific coupling between the electronic states of the neutral D$_2$ and the anionic resonances.

Finally, we found good agreement with experimental data available in literature.

In the future works, we plan to extend the present calculations including excitation to other electronic excited states of D$_2$. In particular, for the couplings to the $^{1,3}\Pi$ states. Moreover, we intent to improuve the model at low energies (temperatures) in order to determine rotational transitions.

The complete cross section sets and the corresponding rate constants presented in the present work can be found at the IAEA database, hcdb \cite{IAEAdb} and LXCat database (www.lxcat.net/Laporta) \cite{doi:10.1002/ppap.201600098}.

\section*{Acknowledgements}

The authors thank Dr. A. Howling (Swiss Plasma Center, Lausanne, Switzerland) for careful reading of the manuscript and discussion. D.R.'s work is carried out under the auspices of the ITER scientist fellowship program (ISFN).


\end{document}